\documentclass[preprint]{elsarticle}
\usepackage{amssymb}
\usepackage{amsmath}
\usepackage{lineno}
\usepackage{placeins}
\usepackage{graphicx}

\begin{document}

\begin{frontmatter}

\title{Generalization of the tensor renormalization group approach to 3-D or higher dimensions}

\author{Peiyuan Teng\corref{cor1}}

\address{Department of Physics,
	The Ohio State University, 
	Columbus, Ohio, 43210, USA}


\cortext[cor1]{Corresponding author}
\ead{teng.73@osu.edu}

\begin{abstract}
In this paper, a way of generalizing the  tensor renormalization group(TRG) is proposed. Mathematically, the connection between patterns of tensor renormalization group and the concept of truncation sequence in polytope geometry is discovered. A theoretical contraction framework is therefore proposed. Furthermore, the canonical polyadic decomposition is introduced to tensor network theory. A numerical verification of this method on the 3-D Ising model is carried out.
\end{abstract}


\end{frontmatter}

	\section{Introduction}

	Tensor network model has become a promising method in simulating classical and quantum many body systems. This method represents physical quantities, such as, wave-function, or the exponential of a Hamiltonian, in terms of a multi-indexed tensor. Then  we can calculate, physical observables, or partition functions,  from a network of tensors. After contracting over this network, we can get physical behavior of our many body system. Examples of this approach is the matrix product state (MPS)\cite{fannes1992finitely}\cite{ostlund1995thermodynamic} and projected entangled paired states (PEPS)\cite{verstraete2004renormalization}.

	The density matrix renormalization group (DMRG) \cite{white1992density} is a powerful method for 1-D quantum systems. For systems in dimensions larger than 1, the DMRG algorithm is known to scale exponentially with the system size. The tensor network correspondence of the DMRG, which is MPS, can be generalized to higher dimensions.  Other generalizations such as multi-scale entanglement variational ansatz (MERA)\cite{vidal2007entanglement} are also key aspects of the tensor network theory.
	
	Compared with quantum Monte Carlo, which suffers from the sign problem, tensor network provides us a new way of doing calculations. Direct contraction of a tensor network, however, is not always possible. As a result, finding an organized way to approximate and contract a tensor network is an important aspect of the tensor network method. For example, we can group together some tensors systematically and contract some of our tensors and get a new coarse-grained tensor. The new tensor network shares the same symmetry with the original tensor network. This idea was explored by Levin and Nave\cite{levin2007tensor}. They proposed this method for 2-D classical lattice models and use singular value decomposition (SVD) to do approximations. Their method has a similar spirit with the block spin method\cite{kadanoff1966spin},a and they call this method tensor renormalization group(TRG). It can be generalized to the so-called second renormalization group (SRG)\cite{zhao2010renormalization}, tensor network renormalization(TNR)\cite{TNR}, higher-order singular value decomposition (HOSVD)\cite{xie2012coarse}.  The way of contracting over the tensor network can also be applied to quantum models using the mapping between a d-dimensional quantum system and d+1 dimensional classical system\cite{jiang2008accurate}. Novel decompositions such as rank-1 decompositions was also proposed to tensor network theory\cite{ran2013theory}.
	
	TRG proposed by Levin and Nave is for 2-D classical systems. For higher dimensional system, especially 3-D, calculations had been done, for example, as a variant of DMRG\cite{nishino2001two}, as a new contraction strategy\cite{garcia2013renormalization}, and HOSVD\cite{xie2012coarse}. Among these methods, HOSVD shares some  similarities with TRG. Mathematically, HOSVD uses Tucker decomposition, which is a specific higher dimensional generalization of SVD. But we should notice that when their method is applied to 2-D system, the geometric structure of the contraction is different from TRG. Therefore it is necessary for us to consider the generalization TRG to higher dimensions. 
	
	In this paper, we propose a framework to do contraction systematically on tensor networks in higher dimensions. We also generalized TRG to higher dimensional tensor network. To achieve this goal, we introduce the canonical polyadic decomposition (CPD) into the tensor network method. The concept of tensor rank can also be defined using CPD, therefore, CPD is also called tensor rank decomposition. Our method reduces to 2D-TRG when it is applied to 2-D tensor network, which is a result of the fact that SVD is the 2-D version of CPD. 
	
	We apply our method to a 3-D cubic tensor network. The concept of the dual tensor network is also proposed. To make the contraction process iterate, the tensor network has to go from the original tensor network to its dual tensor network, then dual back. The dual of the dual tensor network has the same geometric structure with the  original tensor network. 
	
	Mathematically, we propose a correspondence between TRG and the concept of truncation sequence in polytope geometry. The tensor network transforms in the same way as the truncation of a polyhedron(polytope in 4-D or higher). And the tensor CPD geometrically correspond to the truncation of the corner of a polyhedron. TRG in 2-D can also be understood in the framework of dual tensor network and truncation sequence.
	
	This paper is organized as follows. Section \ref{2d}  reviews the tensor renormalization group method in 2-D. Section \ref{3d} generalizes the concept of TRG to 3-D, CPD is introduced and details about the renormalization process are discussed. Section \ref{3dn} shows some simulation results about this 3-D tensor renormalization group method. Section \ref{4d} proposes the similar method in higher dimensions. Section \ref{discuss} discusses some applications and problems with this method.
	
	Regarding the terminology, we need to mention some points that may be ambiguous. The word 'truncation' means either the numerical truncation of the singular values of a tensor(matrix) or the geometric truncation of a polytope. The word 'honeycomb' means either the 2-D honeycomb lattice or tessellation of a higher dimensional space.  
	\section{Tensor renormalization approach for 2-D}\label{2d}
	
	\subsection{Classical Ising model and tensor network}
\begin{figure}[!htb]
			\begin{center}
		\includegraphics[width=80mm]{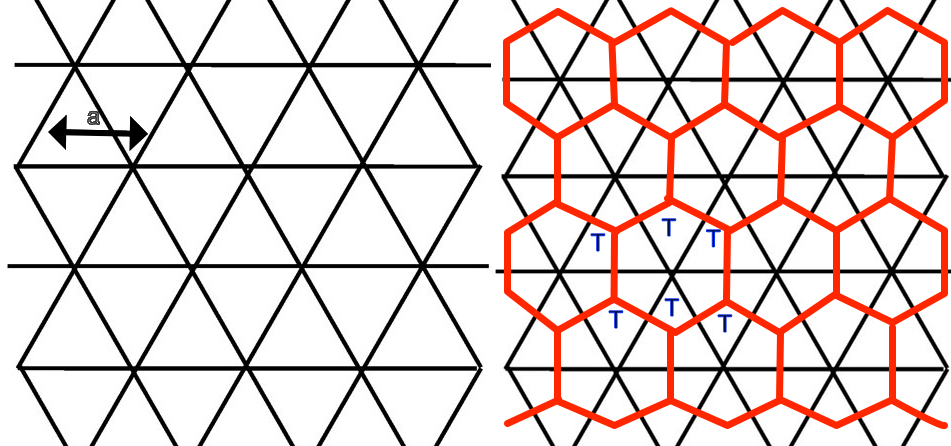}
		\caption{The triangular Ising lattice and its tensor network(red).}
		\label{pictri}
    \end{center} 	
	\end{figure}

\begin{figure}[!htb]
 		\begin{center}
		\includegraphics[width=80mm]{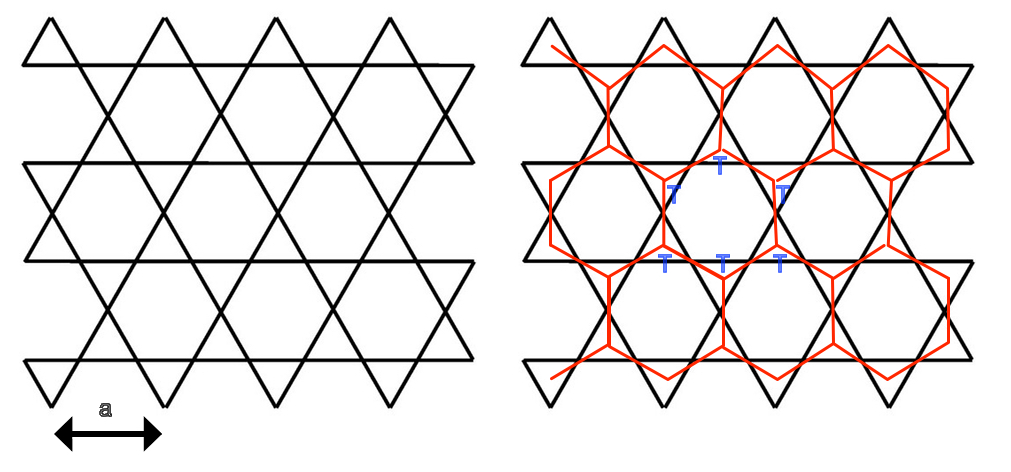}
		\caption{The kagome Ising lattice and its tensor network(red).}
		\label{pickag}
	    \end{center}	
	\end{figure}

	Let's first review 2-D TRG and discuss its geometric meanings. For a classical lattice system, one can find its tensor network representation. For example, both triangular and  kagome lattice can be mapped to a honeycomb tensor network. A honeycomb tensor network may correspond to two Ising model: (1) a triangular lattice Ising model, see Figure \ref{pictri}. (2) a kagome lattice one, see Figure \ref{pickag}. 


	By connecting the centers of the triangles for both lattices, we get a honeycomb tensor network. In these Ising models, the spins live on the vertices and interactions lives on the lines, while for the tensor network, each tensor correspond to a triangle and is represented by $T_{ijk}$. The three indices of the tensor $T_{ijk}$ correspond to three spins of the Ising model.
	
	The partition function of a system can be represented by 
	
	\begin{equation}
		Z=\sum_{spins}e^{-\beta H(\sigma)}=\sum_{indices}T_{ijk}T_{jpq}T_{kab}T_{kmn}...
	\end{equation}

	Here $\beta=(k_b T)^{-1}$, $H(\sigma)$ is the Hamiltonian of the Ising model, and $k_b$ is the Boltzmann constant.
	
	\begin{equation}
		H(\sigma)=-\sum_{<ij>}J\sigma_i\sigma_j-\mu h \sum_{i} \sigma_i
	\end{equation}
	
	For example, for the kagome lattice, we have a factor of $\frac{1}{2}$ in front of $h$, since each spin is shared by two triangles.

	\begin{equation}
		T_{ijk}=e^{\beta J(\sigma_i\sigma_j+\sigma_j\sigma_k+\sigma_k\sigma_i)+\frac{1}{2}\beta \mu h  (\sigma_i+\sigma_j+\sigma_k)}
	\end{equation}
	
	This tensor can be regarded as our initial tensor when specific values of $\beta$, $\mu$ $J$ and $h$ are given. Iterations can be done based on this tensor.
	
	\FloatBarrier 
	\subsection{Contraction of a 2-D honeycomb network }

	The physical properties of a classical spin system can be calculated from the partition function. Direct calculation of the partition function is a very difficult task. Monte Carlo method is an efficient way for classical systems. Here we'd like to use TRG to calculate physical properties.

	First, separate the tensors in the partition function into nearest pairs and find  tensor  $S$ so that
	
	\begin{equation}
		\sum_m T_{ijm}T_{mkl}\approx \sum_n S_{lin}S_{jkn}
	\end{equation}
	
\begin{figure}[!htbp]
		\begin{center}
			\includegraphics[width=60mm]{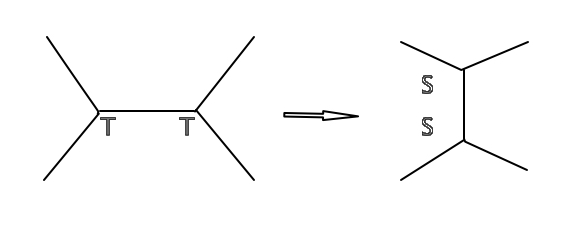}
			\caption{Singular value decomposition(SVD)}
			\label{svd}
		\end{center}
	\end{figure}

	This step can be graphically represented as (shown in Figure \ref{svd})
	
	This step is achieved through singular value decomposition(SVD) by setting 
	
	\begin{equation}
		M_{li,jk}=T_{ijm}T_{mkl}
	\end{equation}
	
	and finding a matrix $S$ which minimizes $||M-S S^T||$. SVD of $M$ gives a diagonal matrix $d$ and two unitary matrices $U$,$V$. 
	\begin{equation}
		M_{li,jk}=\sum_n d_n U_{li,n}V^*_{jk,n}
	\end{equation}
	
	The $S$ matrices can be get by setting $S^A_{lin}=\sqrt{d_n}U_{li,n}$, $S^B_{jkn}=\sqrt{d_n}V^*_{jk,n}$, up to a phase factor. Here we take the largest values of $d_n$ in order to match the dimension of the matrices. Two $S$ matrices can be equal as long as $T$ matrix has certain symmetries and proper unitary phase factor is selected.
	
\begin{figure}[!htbp]
		\begin{center}
			\includegraphics[width=100mm]{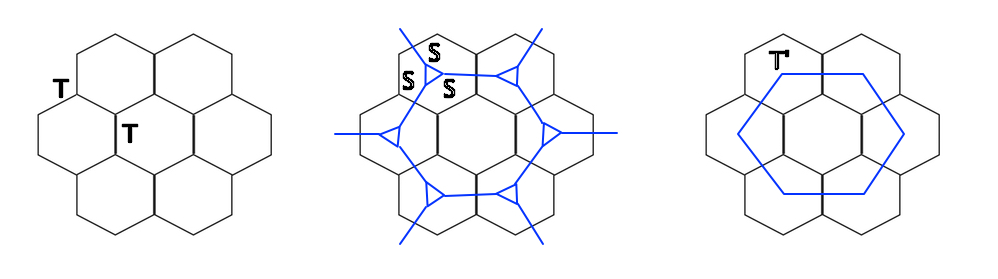}
			\caption{SVD and contraction. Iterating contraction of the honeycomb tensor network. Middle picture is the truncated hexagonal tiling.}
			\label{2dtrg}
		\end{center}
		
	\end{figure}

	Then we contract three $S$ tensors into a tensor $T^{\prime}$, as is shown in Figure \ref{2dtrg}.

	\begin{equation}
		T^{\prime}_{ijk}=\sum_{pqr} S_{iab}S_{jbc}S_{kca}
	\end{equation}
	
	After these steps, the number of spins in a system is reduced by a factor of $\frac{1}{3}$. The tensor is mapped from $T$ to  $T^{\prime}$. When initial temperature and magnetic field is fixed, this method provides us an organized way of approximating and contracting a tensor network.  After several steps of iterations, for a finite system, we'll get a contracted tensor $T^*$. Trace over this tensor, we'll get the value of the partition function. For an infinite system, after proper normalization, this iteration will lead to a tensor $T_{f}$, which is the fixed point of the mapping from $T$ to $T^{\prime}$, although it is not necessarily the critical point tensor of the system. The free energy per spin also converges under iterations.
	
	Here we'd like to introduce some mathematical concepts. The word \textbf{tiling} or \textbf{tessellation}, means using one or more shapes to fill a 2-D plane without overlapping and spacing. A tiling is called \textbf{regular} when the tessellation is done using only one type of regular polygon. There are three regular tilings in 2-D Euclidean plane$\colon$ square tiling, triangular tiling, and hexagonal tiling. The tiling in higher dimensions is usually called a \textbf{honeycomb}. \textbf{Truncation} means cutting the corner of a polygon or a polyhedron. We'll get a new edge or a new face after truncation. The remaining shape depends on the size of the truncation, and this dependence is called a \textbf{truncation sequence}.
	
	From a geometric point of view, a honeycomb lattice can be viewed as a hexagonal tiling of a 2-D plane. The SVD step actually gives rise to a truncated hexagonal tiling of a 2-D plane(see Figure \ref{2dtrg} middle). After summation, we get back the hexagonal tiling. This means a hexagon truncates to itself through a 12-gon(dodecagon). The kagome lattice corresponds to a trihexagonal tiling.

	\FloatBarrier
	
	\subsection{Contraction of a 2-D square tensor network}

	The TRG iteration for a square lattice is similar. In a square lattice we have a tensor with 4 indices, we can write this tensor as $T_{(ij)(kl)}$.
	\begin{equation}
		T_{(ij)(kl)}\approx S_{(ij)r}S_{(kl)r}
	\end{equation}
	
\begin{figure}[!htbp]
		\begin{center}
			\includegraphics[width=60mm]{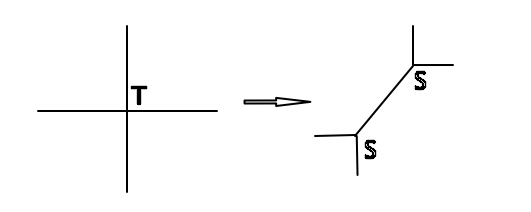}
			\caption{Singular value decomposition(SVD) for a square tensor network. }
			\label{sq}
		\end{center}
		
	\end{figure}

	Treat this tensor as a matrix and do an SVD on this matrix. We can get two tensors denoted by $S$. This step can be graphically represented as (see Figure \ref{sq})
	
\begin{figure}[!htbp]
		\begin{center}
			\includegraphics[width=100mm]{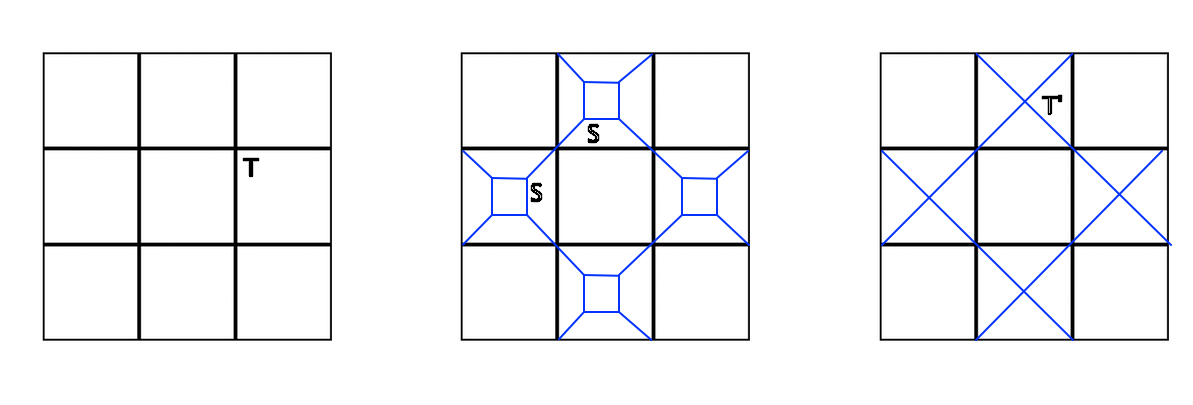}
			\caption{SVD and contraction. Iteration of the square tensor network. Middle picture is the truncated square tiling.}
			\label{sqt}
		\end{center}
		
	\end{figure}

	After SVD, our tensor network now consists of a tessellation of squares and octagons. Summing over every little square, we get our new tensor network $T'$, see Figure \ref{sqt} . This process will iterate and we can therefore calculate the value of the partition function.
	
	We'd like to reconsider TRG from the aspect of tessellation geometry.\textbf{ Square tiling} is another uniform tiling in 2-D. SVD gives rise to a \textbf{truncated square tiling}. The  contraction gives back the square tiling. When thinking in terms of tessellation, we discover that $T'$ is actually the dual tensor network of the square tensor network $T$. The tricky part is that, for 2-D, the dual of a square is still a square, therefore, we may not discover the subtleties in 2-D. Things are not so simple in higher dimensions, since the dual of a cube is no longer a cube. Our framework of higher dimensional TRG reduces to the square case when applied to a 2-D system. We'll explain the concept of a dual tensor network in details later.

\begin{figure}[!htbp]
		\begin{center}
			\includegraphics[width=120mm]{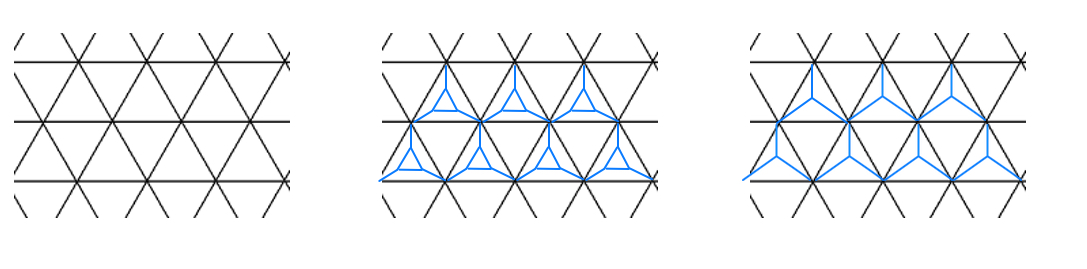}
			\caption{Converting a triangular tensor network (black) to a hexagonal tensor network (blue). Tucker decomposition and contraction.}
			\label{tritucker}
		\end{center}
		
	\end{figure}

	\textbf{Triangular tiling} is the third type of tiling which consists of only one shape. A triangular tensor network can be converted to a hexagonal tensor network using Tucker decomposition, see Figure \ref{tritucker}. Therefore, we have discussed the contraction technique for all the possible regular tensor network in 2-D for completeness.

	\subsection{Free energy calculation for a 2d kagome Ising model}
	
	The TRG contraction technique can be applied to both the finite lattice and the infinite lattice. Applications to a finite lattice system have been discussed in earlier papers, for example, HOSVD. In this part we'd like to use TRG to calculate the physical properties of an infinite lattice.
	
	Now let's discuss an example of how to calculate the physical properties as a function of the temperature and the magnetic field. For simplicity, we assume that the spin takes value -1 or 1, so each index of tensor $T_{ijk}$ can have two values. Then this tensor can have 8 variables. For our kagome Ising lattice the Hamiltonian is symmetric about the permutation of $a$, $b$, and $c$. So there are actually 4 variables and the matrix $M$ is symmetric, so tensor $S^A$ and $S^B$ are the same.

	Following the steps mentioned above, i.e. SVD and contraction, we can get  the tensor $T^{\prime}$. This tensor actually contains the information of 3 tensors in the previous step.  After several iterations, this tensor will get larger and larger. For an infinite system, it will diverge. To avoid divergence, we should normalize this tensor after each iteration.  One of the choices of the normalization factor is the largest value of $\sqrt{d_n}$. There are some subtle parts about the selection of this scaling factor, we will discuss it in detail in the 3-D case, see section \ref{marker43}. Similar normalization factor had been noticed by\cite{li2011linearized}.

	To calculate the partition function, we assume this system have $N$ tensors. Because of the geometry of a kagome lattice, $N$ tensors correspond to $1.5N$ spins.
	\begin{equation}
		Z=\sum_{spins}e^{-\beta H(\sigma)}=\sum_{indices}T_{ijk}T_{jpq}T_{kab}T_{kmn}...
	\end{equation}
	
	Next step is to separate these $N$ tensors into pairs and do SVD and contraction. The details of this part have been discussed previously.
	
	Let's denote the original tensor by $T_0$ and the contracted tensor by $T_1$. Due to the translational symmetry of the tensor network, this tensor is the same everywhere in the tensor network. Furthermore, let's the normalization factor by $f_1$. We have $N$ tensors, so the total normalization factor should be $f_1^{N/3}$, so after one step of iteration
	
	\begin{equation}
		Z(\beta,h)=\sum \prod_N T_0(\beta,h)=f_1^{N/3}  \sum \prod_{\frac{N}{3}} T_1(\beta,h) 
	\end{equation}  
	
	Denote the normalization factor of the second iteration by $f_2$, and continue this iteration, we can get
	
	\begin{multline}
		Z(\beta,h)=\sum \prod_N T_0(\beta,h)=f_1^{N/3} \sum \prod_{\frac{N}{3}} T_1(\beta,h)\\=f_1^{N/3} f_2^{\frac{N}{3^2}}   \sum \prod_{\frac{N}{3^2}} T_2(\beta,h)=... 
	\end{multline} 
	
	In the end,  we'll get
	
	\begin{equation}
		Z(\beta,h)=f_1^{N/3} f_2^{\frac{N}{3^2}}...f_n^{\frac{N}{3^{n}}}.....Tr(T_f)
	\end{equation} 
	
	 $T_f$ is the fixed point of the mapping from $T$ to $T'$. The free energy can be calculated as
	
	\begin{multline}
		F(\beta,h)=-\frac{1}{(3\times 1.5N)\beta}ln Z(\beta,h)\\=-\frac{1}{(4.5)\beta}(ln f_1+\frac{1}{3}ln f_2+....\frac{1}{3^{n-1}}ln f_n+...+\frac{1}{N} Tr(T_f))
	\end{multline}
	
	The factor of $1.5$ comes from the kagome lattice. If  $N$ goes to infinity,
	
	\begin{multline}
		F(\beta,h)=-\frac{1}{(3\times 1.5N)\beta}ln Z(\beta,h)\\=-\frac{1}{(4.5)\beta}(ln f_1+\frac{1}{3}ln f_2+....\frac{1}{3^{n-1}}ln f_n+...)
	\end{multline}

	These factors are also a function of temperature and magnetic field. One important thing is that, for a infinite lattice, i.e. when N goes to infinity, this factor has to be carefully chosen so that the limiting value of $f$ neither converges to zero, nor diverges. We assume this limits exists and the limiting $f$ is an none zero value at almost every point of the parameter space of $t$ and $h$, except for the critical point.
	
	Numerically, we could evaluate the free energy with and without a small magnetic field. By taking the numerical differentiation with respect to $h$, we can get the magnetization as a function of $T$.

	\section{Tensor renormalization approach for 3-D}\label{3d}
	\subsection{Dual polyhedron}
	In this section, we consider the problem of how to contract a 3-D cubic tensor network under tensor RG. The concept of \textbf{dual polyhedron}\cite{wenninger2003dual} can be used to illustrate this process.
	
\begin{figure}[!htbp]
		\begin{center}
			\includegraphics[width=40mm]{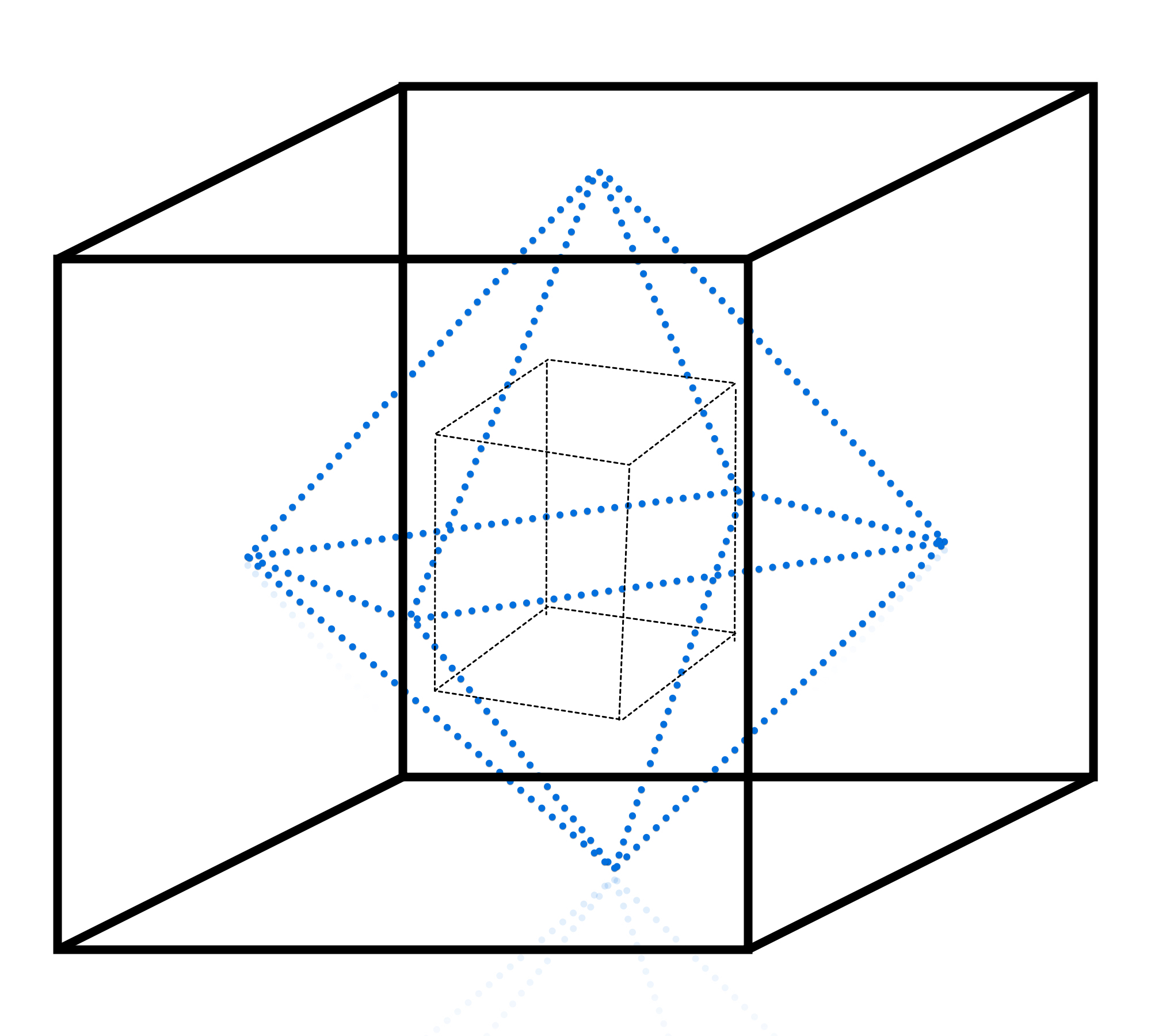}
			\caption{Dual polyhedron for a cube. The dual polyhedron of a cube is a octahedron. The dual polyhedron of a octahedron is a cube.}
			\label{fdual}
		\end{center}
	\end{figure}

	For some polyhedrons in 3 dimension, the dual polyhedron can be defined by converting each face into a vertex and each vertex into a face. Therefore, for a cube, the dual polyhedron will be a octahedron. Again by converting each face of the octahedron into a vertex, we will get a cube. This process is illustrated in Figure \ref{fdual}. The dual of a dual polyhedron is the polyhedron itself.
	
	\subsection{Tensor renormalization group (TRG)}
		
\begin{figure}[!htbp]
		\begin{center}
			\includegraphics[width=80mm]{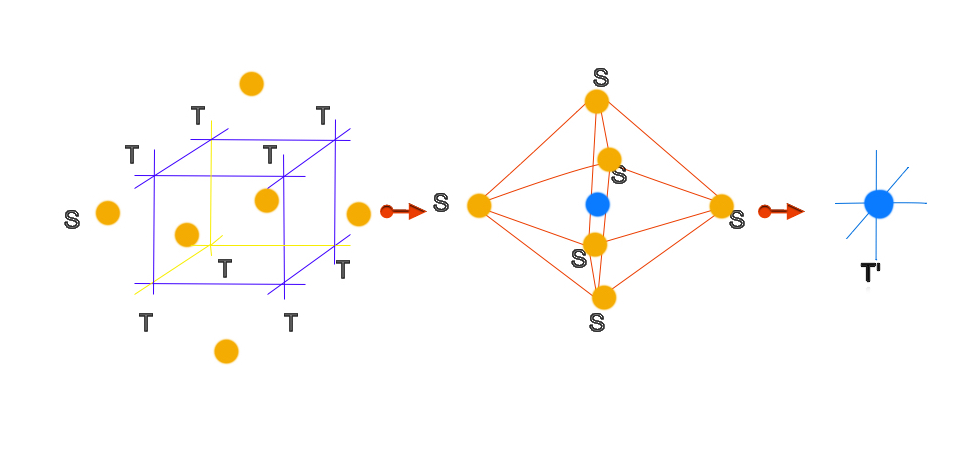}
			\caption{Graphical representation of the Tensor RG process, from $T$ to $T\prime $. We start from a cubic tensor network $T$. Eight $T$ tensors forms a cube and this cube has 24 external tensor legs. We can view our space as a stacking of this tensor cube. Our first step is to change from the $T$ tensor network to the $S$ tensor network(we have marked the relative positions between $T$ and $S$. Same type of tensor is marked with the same color.). The external legs of the $T$ tensors forms part of a cube which surrounds tensor $S$, and this cube is contracted to cube $S$. The second step is to contract 6 $S$ tensors to construct a new tensor $T'$. Detailed implementation of these two steps are explained in section \ref{3p4} and\ref{3p5}.}
			\label{3trg}
		\end{center}
	\end{figure}

	Let apply the concept of dual polyhedron to our cubic tensor network. For the first step, we can separate the cubic tensor network by grouping together 8 tensors of one cube and transform it into an octahedron of 6 tensors. For the second step, for each tensor of the octahedron, we can do an SVD, then sum over the octahedron, we'll get our renormalized tensor $ T'$. This process is shown in Figure \ref{3trg}.

We start with a partition function which can be represented by a cubic tensor network. 
	\begin{equation}
		Z_{Ising}=\sum_{indices} T_{abcxyz}T_{ab'c'x'y'z'}T_{aâbc'x'y'z'}...
	\end{equation}
	
	The goal is to find an iterative way to simplify this partition function and write it in terms of $T'$.
	
	\begin{equation}
		Z_{Ising}=\sum_{indices} T'_{abcxyz}T'_{ab'c'x'y'z'}T'_{abc'x'y'z'}...
	\end{equation}
	
	\subsection{Canonical polyadic decomposition (CPD)}	
\begin{figure}[!htbp]
		\begin{center}
			\includegraphics[width=80mm]{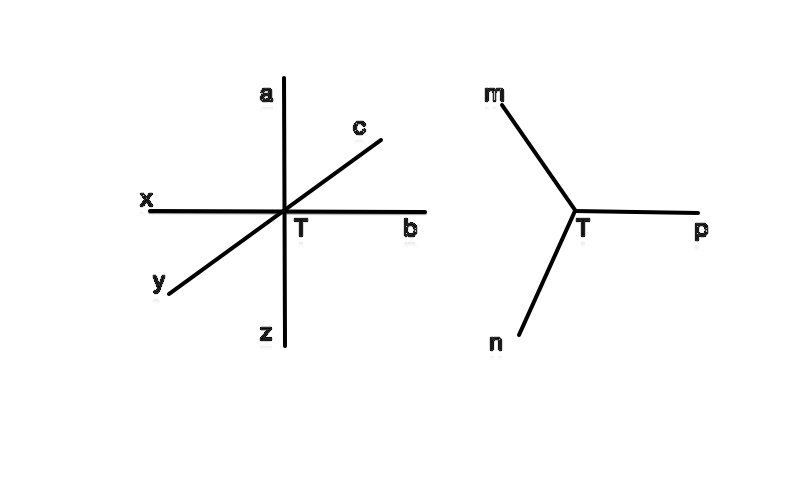}
			\caption{Relabel $T_{abcxyz}$ and get $T_{mnp}$, $(ax)\leftrightarrow m$,$(by)\leftrightarrow n$,$(cz)\leftrightarrow p$}
			\label{10}
		\end{center}
	\end{figure}

In our tensor network, each tensor can be written as $T_{abcxyz}$, which have $D^6$ elements, where $D$ is the bond dimension. Our starting element of TRG is a cube which have 8 tensors. For each tensor we label outgoing indices(with respect to the cube) and by $abc$ and the internal indices are labelled by $xyz$. Therefore we can change the tensor from a tensor with 6 indices $T_{abcxyz}$ to a tensor with 3 indices $T_{(ax)(by)(cz)}=T_{mnp}$, see Figure \ref{10}, where $(ax)$ is treated as one index . This tensor have $(D\times D)^3$ elements. Notice that we are just relabeling the tensors in order to do decompositions, so $ax$ and $m$ are equivalent.	

	Now we introduce a new way of transforming  tensors in tensor network theory. This method is called "\textbf{tensor rank decomposition}" or "\textbf{canonical polyadic decomposition (CPD)}" We call it CPD for short in this article. It can be regarded as a tensor generalization of the widely used singular value decomposition (SVD).
	
	For a three-way tensor, CPD can be written as
	
	\begin{equation}
		T_{mnp}=\sum_{r}\lambda_r A_{rm}\otimes A_{rn}\otimes A_{rp}
	\end{equation}
	
	This can be regarded as a three dimensional generalization of the SVD, which, for a matrix, is
	\begin{equation}
		T_{mn}=\sum_{r}\lambda_r A_{rm}\otimes A_{rn}=U S V^{*}
	\end{equation}
	
 $S$ is the singular value matrix and the singular values are $\lambda_r$, $U$ and $V$ correspond to $A_{rm}$ and $A_{rn}$.
		
	The \textbf{rank} of a n-way tensor can be defined as the minimal value of r where the following expression is exact.
	
	\begin{equation}
		T_{mnp...z}=\sum_{r}\lambda_r A_{rm}\otimes A_{rn}\otimes A_{rp}\cdots\otimes A_{rz}
	\end{equation}
	
	We can do a minimal square fitting for any integer value of $r$. Rank correspond to the minimal $r$ when this composition is exact. When $r$ is larger than the rank of the tensor, we may have multiple solutions. When $r$ is smaller that the rank, we can fix $r$ and find the least square approximation by minimizing $||T-M||^2$.
	
	\begin{equation}
		T_{mnp...z}\approx \sum_{r}\lambda_r A_{rm}\otimes A_{rn}\otimes A_{rp}\cdots\otimes A_{rz}=M
	\end{equation}
	
	A detailed review of tensor decomposition could be find in\cite{kolda2009tensor}. Currently, there is no good way to find the rank of an arbitrary tensor\cite{kolda2009tensor}. For a three way tensor, however, an upper and lower bound can be set as\cite{kruskal1989rank}
	
	\begin{equation}
		max(I,J,K)\leq rank(T_{mnp})\leq min(IJ,JK,KI)
	\end{equation}
	
	Here $T_{mnp}$ is a $I\times J\times K$ array. For example, for a $4\times 4\times 4$ tensor T, which is the tensor that we'll use in our calculation, the rank of this tensor should be between 4 and 16.
	
	We should also notice the difference between CPD and higher order singular value decomposition(HOSVD).
	There are two major ways to decompose a tensor, one is CPD, the other is HOSVD (Tucker decomposition). These two techniques are already used in subjects like signal processing etc. Compared with CPD, HOSVD can be written as
	
	\begin{equation}
		T_{mnp...z}=\sum_{r}\lambda_{\alpha\beta\gamma...\omega} A_{\alpha m}\otimes A_{\beta n}\otimes A_{\gamma p}\cdots\otimes A_{\omega z}
	\end{equation}
	
	We can see in CPD, $\lambda$ are numbers while in HOSVD  $\lambda$ is a tensor. CPD and HOSVD can be regarded as two different ways of generalizing matrix SVD. The difference is how to understand the singular values (numbers or matrices) in a SVD. Applications of HOSVD to Ising models can be found in\cite{xie2012coarse}.
	
	\subsection{TRG in detail: From cube to octahedron}
	\label{3p4}

\begin{figure}[!htbp]
			\begin{center}
				\includegraphics[width=100mm]{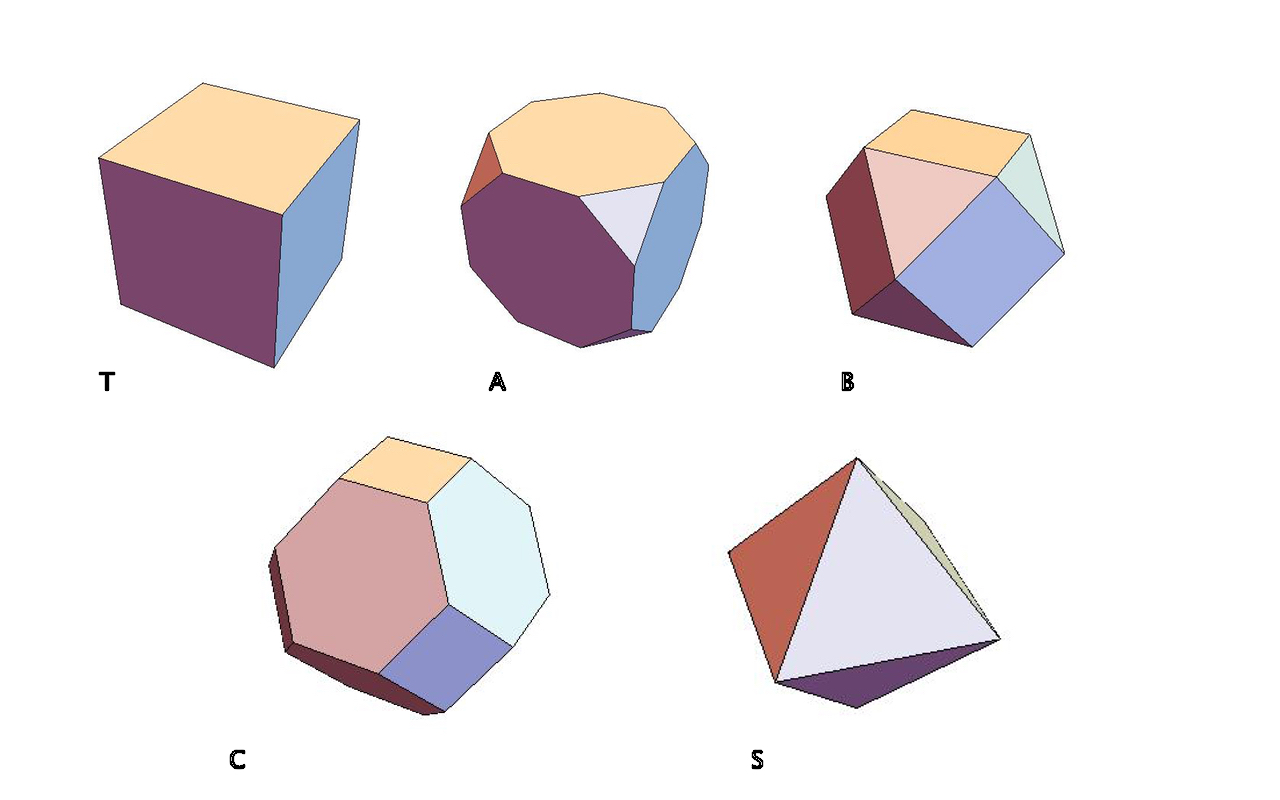}
				\caption{Truncation sequence from a cube to an octahedron. Letters represent the corresponding tensor network. Edges correspond to the connections in the tensor network. Vertices correspond to the places where the tensors live. Each face have 4 perpendicular external legs (except for $S$), these legs are omitted in the picture.}
				\label{tabc}
			\end{center}
		\end{figure}

	The idea of this process can be get from the process of changing a cube to an octahedron shown in Figure \ref{tabc}. We can think of the tensor network as the stacking of these basic elements. The letters represent different types of tensor network that will be discussed later. External legs are omitted.

	To continue TRG, we may start from the three-way tensor constructed from the Ising model, and apply a CPD on this tensor. We can choose the number of singular values to be a number $k$ and make a least square fitting. 
	
	\begin{equation}
		T_{mnp}\approx \sum_{r}\lambda_r A_{rm}\otimes A_{rn}\otimes A_{rp}
	\end{equation}

	The number of $\lambda_r$ to keep should be determined by balancing the computational cost and accuracy. The expression above is exact when it is summed to the rank of the tensor.
	
	The CPD is not unique in general, under some conditions it can be unique up to rescaling and change of basis. Detailed discussion of the uniqueness of CPD can be found in\cite{kruskal1989rank}.
	
	We can furthermore absorb $\lambda$ in to $A$ matrices by multiplying the cubic root of $\lambda_r$ into each $A_r$,$A'_{rm}=\sqrt[3]{\lambda_r}A_{rm}$, so we can write
	
	\begin{equation}
		T_{mnp}\approx \sum_{r} A'_{rm}\otimes A'_{rn}\otimes A'_{rp}
	\end{equation}
	
	According to the definition of the outer product $'\otimes'$ of a tensor, for each specific $m,n,p$. The value of $T_{mnp}$ should just take the numerical product of $A'_{rm}$ and $A'_{rn}$ and $A'_{rp}$.
	
\begin{figure}[!htbp]
		\begin{center}
			\includegraphics[width=80mm]{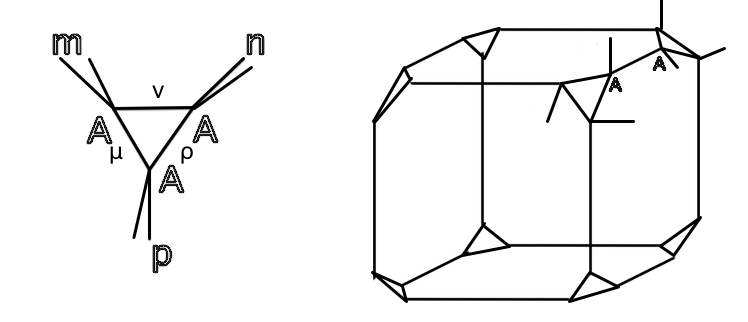}
			\caption{ CPD changes our network from $T$ tensors to $A$ tensors, some external tensor legs are omitted for simplicity, although they are there.}
			\label{ta}
		\end{center}
	\end{figure}

	In order to get rid of the tensor product symbol and write it in the formalism of tensor network, we can convert the previous equation into, see Figure \ref{ta}.
	
	\begin{equation}
		T_{mnp}\approx \sum A'_{\mu\nu m} A'_{\nu \rho n} A'_{\rho \mu p}
		\label{25}
	\end{equation} 
	
	where tensor $A'$ is not zero only when $\mu=\nu=\rho$. This expression is identical with the previous one, although we can have a straight forward graphical correspondence for this expression. We have to point out that a CPD may not be symmetrical, therefore we may need to keep track of three different $A'$ tensors inside a cube.
	
	Therefore our tensor networks can be written in terms of $A'$, this process is shown in Figure \ref{ta}.
	
	\begin{equation}
		Z\approx \sum A'\cdots A'
	\end{equation}

\begin{figure}[!htbp]
			\begin{center}
				\includegraphics[width=80mm]{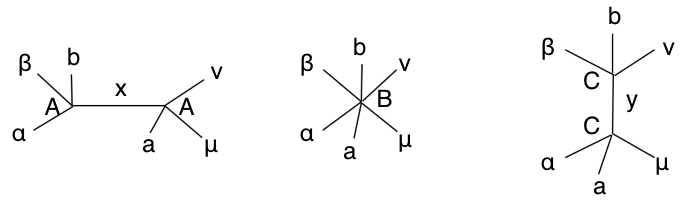}
				\caption{ The contraction from SVD from $A$ to $B$ to $C$}
				\label{abc}
			\end{center}
		\end{figure}

	These $A'$ matrices can be paired up by summing over every bonds on the edges of the cubes, see Figure \ref{abc}.   We can label these summed matrices(tensors) as $B$ and 
	
	\begin{equation}
		B_{\alpha\beta a\mu\nu b}=\sum_x A'_{\alpha\mu ax} A'_{\beta\nu bx}
	\end{equation}
	
\begin{figure}[!htbp]
		\begin{center}
			\includegraphics[width=100mm]{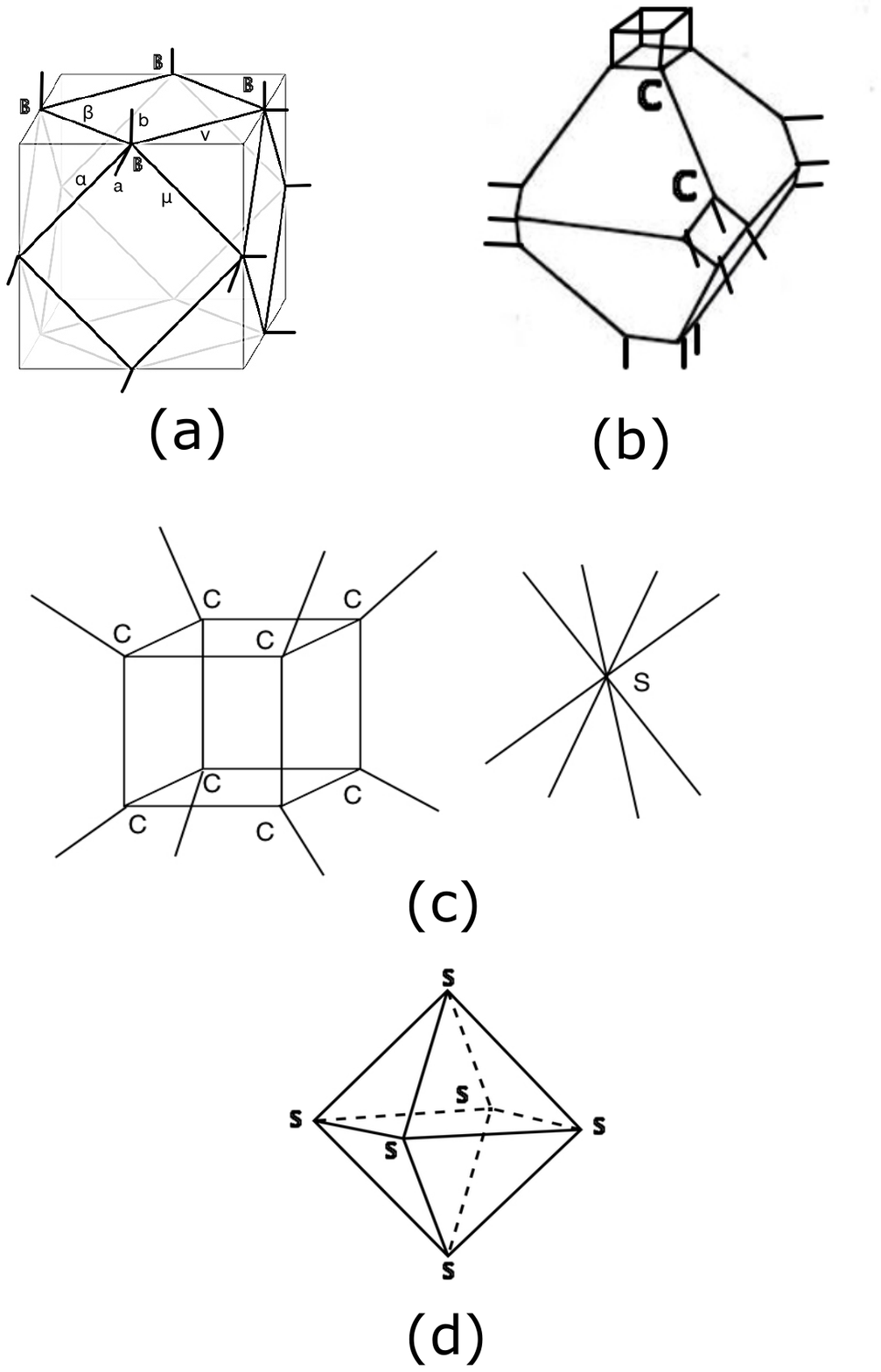}
			\caption{(a) CPD changes our network from $T$ tensors to $A$ tensors, some external tensor legs are omitted for simplicity, although they are there.    (b) $B$ tensor network (cuboctahedron). The $T$ network is drawn in thin lines for reference, some lines are omitted for simplicity. (c) Contraction over 8 $C$ tensors and $S$. (d) $S$ tensor network. $S$ tensors lives on octahedrons. This is the dual tensor network of the cubic $T$ tensor network. Each $S$ tensor have 8 legs, but we draw 4 of them for clearness.}
			\label{drawing}
		\end{center}
	\end{figure}

	These $B$ tensors form a tensor network of \textbf{cuboctahedron}, which have 8 triangular faces and 6 square faces, see Figure \ref{drawing} (a).
   
	\begin{equation}
		B_{\alpha\beta a\mu\nu b}=B_{kl}=\sum_y C_{\alpha\beta a y}C_{\mu\nu b y}=\sum_x A'_{\alpha\mu ax} A'_{\beta\nu bx}
	\end{equation}
	
 This $B$ tensor can be written in terms of a matrix $B_{kl}$ by grouping together indices.  Then we can do an SVD (see Figure \ref{abc}) and separate it into two different matrices and label them by $C$, see Figure \ref{drawing} (b).
	
	\begin{equation}
		Z\approx \sum C\cdots C
	\end{equation}
	
	In the expressions above, 8 $C$ matrices that forms a cube can be grouped together and summed over and therefore we get another matrix and it is labeled by $S$, see Figure \ref{drawing} (c).

	\begin{equation}
		S=\sum_{cube}CCCCCCCC
	\end{equation}
	
	Notice these $S$ matrices have 8 indices. The number of values in each index depends on the cut-off imposed on the singular matrix in our previous SVD.
	
	Now we have done a transformation from a cubic tensor network $T$ to octahedron tensor network $S$, see Figure \ref{drawing} (d). Notice the length of the edge of the octahedron is $\sqrt{2}$ times larger than that of the cube.
	
	\begin{equation}
		Z\approx \sum 
		S\cdots S
	\end{equation}

\begin{figure}[!htbp]
		\begin{center}
			\includegraphics[width=40mm]{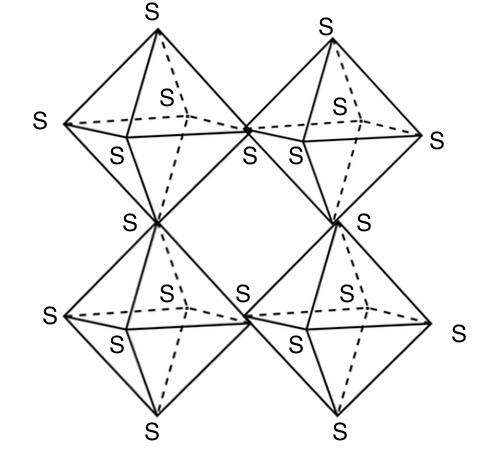}
			\caption{Stacking of the octahedrons in the dual space.}
			\label{sstack}
		\end{center}
	\end{figure}

	In this dual tensor network, each tensor S have 8 legs. The bond dimension in each leg is determined by the SVD in matrix $B$. Each tensor S are connected 8 other tensors. Figure \ref{sstack} shows how these octahedrons are stacked together.
	
\begin{figure}[!htbp]
		\begin{center}
			\includegraphics[width=40mm]{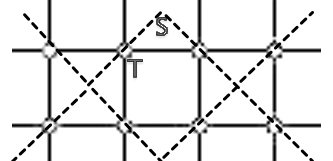}
			\caption{The projection of 3-D model to 2-D model}
			\label{proj}
		\end{center}
	\end{figure}

	We'd like to point out that this 3-D cubic tensor renormalization group approach could be projected to a 2-D plane that parallels to the faces of the 3-D network. The patterns of this projection exactly correspond to the renormalization approach to a 2-D square tensor network proposed in \cite{levin2007tensor}. This is the reason that we say our framework is a generalization of TRG. This correspondence is shown in Figure \ref{proj}. The easy thing for a 2-D square network is that the dual of a square is still a square. The orientation of the dual square is changed, but when being converted back, the dual of dual square have the same orientation as the original one.
	
	We should give a short mathematical background review of this process. The process shown above, in the aspect of polyhedron geometry, is called the \textbf{truncation sequence} of a polyhedron. Our sequence starts from a cube, then goes through truncatedcube, cuboctahedron, truncatedoctaheron, to octahedron.  The cuboctahedron is called rectified or complete-tuncation. The cube and the octahedron are two of the 5 Platonic solids (convex regular polyhedron). The truncatedcube, cuboctahedron, truncatedoctaheron are part of the so-called 
	Archimedean solid when the lengths of its edges are equal.
	
	For 3-D Platonic solids another two sequences exists. Icosahedron and dodecahdron truncates to each other. A tetrahedron rectifies to a octahedron and truncates to itself. 
	
	\subsection{TRG in detail: From octahedron back to a cube}
	\label{3p5}
\begin{figure}[!htbp]
		\begin{center}
			\includegraphics[width=50mm]{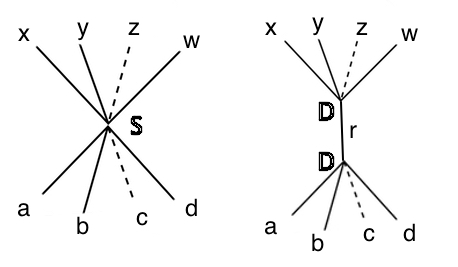}
			\caption{SVD of the dual tensor network $S$.}
			\label{sd}
		\end{center}
	\end{figure}

	The next step is to transform back  to the cubic tensor network from the dual space (octahedron tensor network). This step is straight forward, and it involves with only one step of SVD. We separate 8 indices of the $S$ tensor network into 2 parts and treat each part as a single index. Then $S$ becomes a matrix and we can do an SVD on it, see Figure \ref{sd}. 
	
	We should impose a cut-off on singular values. In order to make the renormalization process iterate, the bond dimension should stay the same. This is a requirement of renormalization.

	\begin{equation}
		S_{(abcd)(xyzw)}=\sum_r D_{(xyzw)r}D_{(abcd)r}
	\end{equation}

\begin{figure}[!htbp]
		\begin{center}
			\includegraphics[width=60mm]{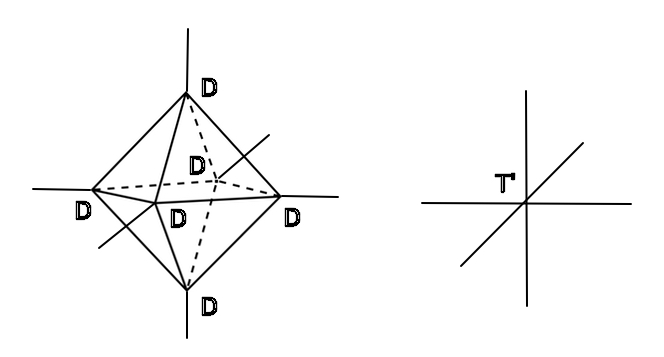}
			\caption{The renormalized tensor network $T'$}
			\label{tp}
		\end{center}
	\end{figure}

	The last step is summing over the $D$ tensors. 6 $D$ tensors live on a octahedron and they can be summed over to get a renormalized tensor $T'$,see Figure \ref{tp}. Finally, we can express our partition function in terms of $T'$.

	\begin{equation}
		Z_{Ising}=\sum_{indices} T'_{abcxyz}T'_{ab'c'x'y'z'}T'_{abc'x'y'z'}...
	\end{equation}
	
	Due to the summation, the norm of this tensor is getting larger. In order to restore the original tensor and make this system scale invariant, we need to rescale this tensor by dividing some factors.  In fact, for a 3-D model, we could do a rescale on both the tensor space and its dual space. Details of this rescale will be discussed later.
	
	In summary, the renormalization of a cubic 3-D tensor network can be realized by CPD, SVD and contraction. We are writing our partition function in terms of $T$-type tensors, $A$-type, $B$-type, $C$-type, $S$-type, $D$-type, and $T'$($T$)-type tensors. Rescaling should be done on both $S$-type and $T$-type tensor, i.e. the original space and the dual space. 
	
	\section{Numerical results of the 3-D cubic tensor network model}\label{3dn}
		\subsection{Ising model and tensor network}
		 
\begin{figure}[!htbp]
			\begin{center}
				\includegraphics[width=80mm]{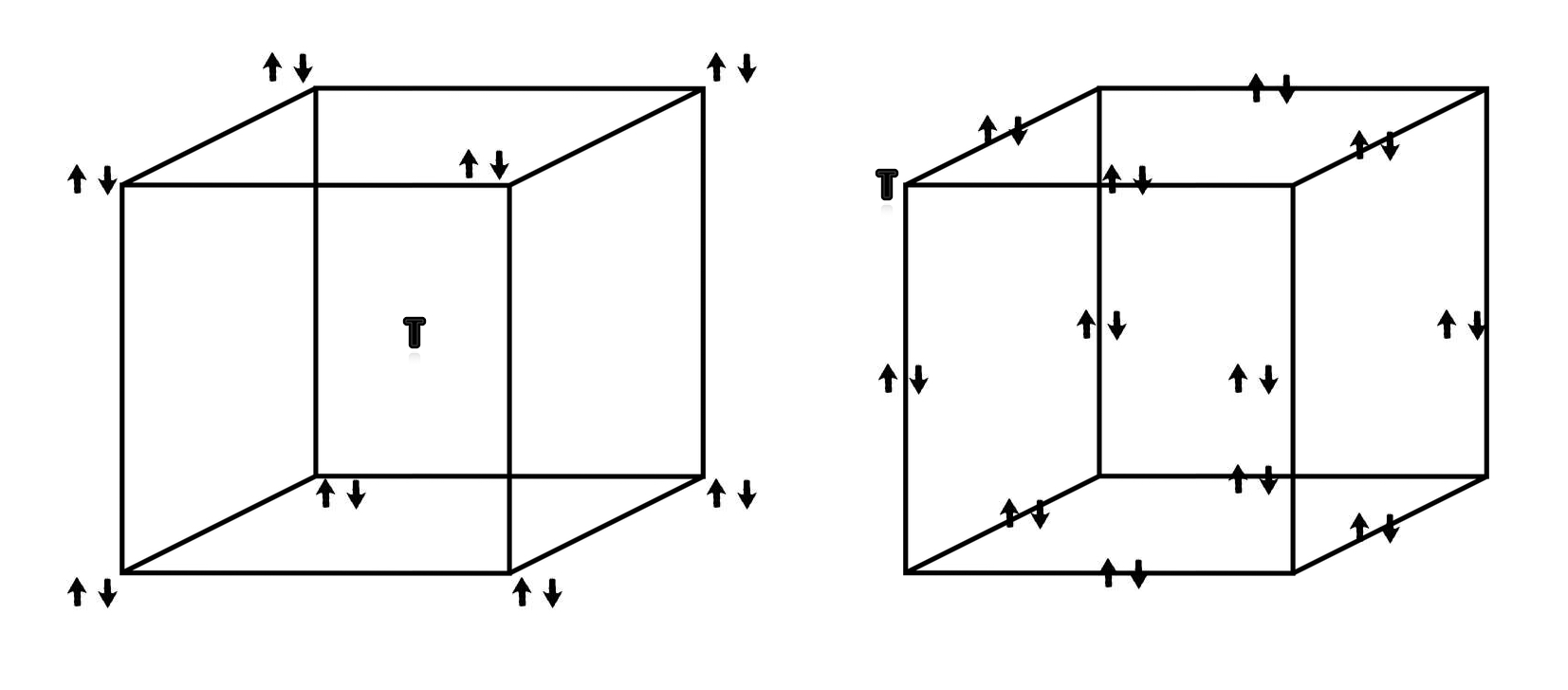}
				\caption{Vertex Ising model and Bond Ising model}
				\label{vb}
			\end{center}
		\end{figure}

		 We'd like to test our framework by considering Ising model with spins that live on bonds. Both the cubic Ising model with spins that live on vertices and the Ising model spins that live on bonds can be represented by a 3-D cubic tensor network. We call them \textbf{vertex Ising model} and \textbf{bond Ising model} respectively, see Figure \ref{vb}. A bond Ising model requires less tensor bond dimension. For example, in $Na_3N$, the $Na$ atom have a natural stacking as a 3-D bond Ising model. Therefore it is also practical to discuss the bond Ising model.
		
	\FloatBarrier	
	\subsection{Calculation framework}
	
	We realize our tensor network model using MATLAB.  The SVD function is embedded in MATLAB. The CPD calculation is done using the MATLAB Tensor Toolbox Version 2.6\cite{bader2012matlab} developed by Bader and Kolda et al. CPD had been developed as a workable MATLAB code due to its widely application to multilinear problems in chemometrics, signal process, neuroscience and web analysis\cite{acar2011scalable}. Different algorithms have been provided by this package, such as Alternation Least Square(ALS) optimization and  Gradient-based optimization. methods\cite{acar2011scalable}. In our code, we use Alternation Least Square$(cp\_als)$. 
	
	Based our description, our code consists of 6 types of tensors, $T$, $A$, $B$, $C$, $S$, $D$, $T'$. They are related by CPD, contraction, SVD, contraction, SVD, contraction, respectively. For the contraction part, we need to contract over a bond, a cube and an octahedron respectively.

	\subsection{Calculation steps: tensor size and cutoff}
	
	First, we need to set up our initial tensor from the Hamiltonian. Our initial tensor is based on an Ising model with spins living on bonds. For each vertex, the spins form a octahedron. We only consider the nearest neighbor interaction, the interaction terms can be represented by the edges of the octahedron. 
	
	The initial tensor can be represented as a tensor with 6 indices.
	
	\begin{equation}
		T_{0}(h,T)=e^{\beta J(\sum_{ij}a_i a_j)+\frac{1}{2}\beta \mu h  \sum_{i=1}^6 a_{i}}
	\end{equation}
	
	We may choose $\mu=1$, $J=1$, then our initial tensor is a function of temperature, and the magnetic field. 
	
	To be specific, we start from a model with spins taking two values, $\pm 1$. Therefore our initial tensor is $2\times2\times2\times2\times2\times2$. We  group together 2 indices, then our tensor is $T_{ijk}$ has a dimension of $4\times4\times4$. After setting up the initial tensor,  we can do CPD on this tensor. For the reason of computational time and details of the algorithm, we choose only the largest value of CPD. It is obvious that this choice will not work for the zero magnetic field case, due to the form of the tensor at zero magnetic field, which has a $Z_2$ symmetry.
	
	In our calculation, we take the largest value for CPD, and 2 largest singular values for SVD.  Cut-off is imposed for the purpose of renormalization. Our cut-off stress the speed of the iteration and our calculation is done on a personal computer.
	
	Section \ref{marker43} and section \ref{marker44} is about the rescaling and free energy calculation. Readers who are not interested can skip them.
\subsection{Accuracy of CPD}

In this section, we'd like to access the accuracy of CPD. A test is done on a $4\times4\times4$ tensor, which is the starting point of our iteration.

To achieve this goal, we  verify CPD, by comparing the value of the trace of a single tensor with our without CPD (direct calculation) to see how much error can be brought by CPD. We plot the error of CPD as a function of $r$, i.e. how many spectrum values are kept in CPD.

\begin{equation}
T_{mnp}=\sum_{r}\lambda_r A_{rm}\otimes A_{rn}\otimes A_{rp}
\end{equation}

\begin{figure}
\begin{flushright}

		\includegraphics[width=130mm,trim=10mm 0mm 0mm 0mm]{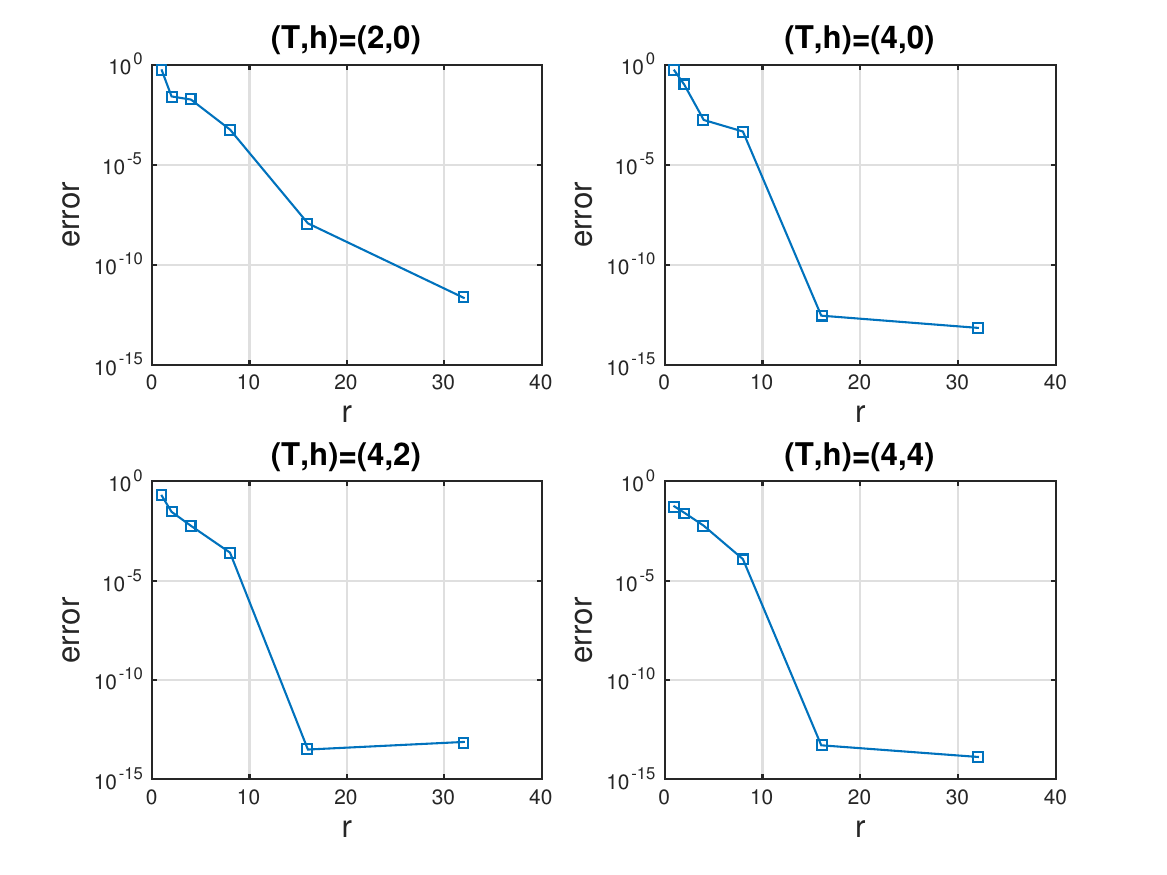}
		
		\caption{Semi-log diagram of the percentage error  as a function of $r$ under different magnetic field and temperature using $cp\_als$. The $y$ axis is the log of the percentage error  $|(Z-Z_0)/Z|$ and $x$ axis is the log of $r$. $r=1,2,4,8,16,32$.  $(T,h)=(2,0);(4,0);(4,2);(4,4)$ respectively. The error decreases with the increase of $r$. We notice that the error may increase when $r$ is larger that the theoretical rank of the tensor. The fitting will still be accurate since it is around the numerical truncation error. }
		\label{acpd}
\end{flushright}	
\end{figure}

We estimate the error under some combinations of $T$ and $h$, see Figure \ref{acpd}. For a nonzero $h$, the lowest order could have a truncation error of about $10\%$. 

	\subsection{Rescaling of the tensor network and the dual tensor network}
    \label{marker43}
	For the reason of simplicity, we'd like to use another notation to describe the asymptotic behavior. The observation is that our Hamiltonian is asymptotically scale invariant.
	
	After proper cut-off is set on the tensors, similar to the 2-D case, we still need to rescale the tensors. We should notice that our new tensor $T'$, after one step of RG, contains the information of 8 tensor $T$. Therefore if we do not do a rescale, the tensor $T$ will blow up.  
	
	We'll use the argument from dilation invariance to find the scaling factor. Under dilation, the Hamiltonian transforms as
	\begin{equation}
		H(\lambda \beta)\sim \lambda H(\beta)
	\end{equation}

	As a result, for the tensor, we have
	\begin{equation}
		||T(\lambda \beta)||\sim ||T(\beta)||^{\lambda}
	\end{equation}

	For a 2-D honeycomb tensor network, we have a contraction over a triangle $T'_s=TTT$, we use the subscript $s$ to denote the tensor before scaling. Then
	
	\begin{equation}
		||T'_s(\beta)||\sim ||T( \beta)||||T( \beta)||||T( \beta)||
	\end{equation}
	
	In order for our new tensor $T'$ to have the same dilation relation, we need to divide it by a factor $f$, where $f \sim T^2$, which have the same order with the largest singular value in SVD.
	\begin{equation}
		||T'(\beta)||\sim ||T'_s(\beta)||/f
	\end{equation}
	
	Then we'll have
	
	\begin{equation}
		||T'(\lambda \beta)||\sim ||(T'(\beta))||^{\lambda}
	\end{equation}
	
	If we don't impose a rescaling, the tensor gets bigger and bigger. A natural way to select this factor would be the largest singular value of SVD.  The reason for us to do the rescaling is that we need to keep iterating, and $T'$ and $T$ should be physically equivalent. The scaling factor is extracted and it contributes to the free energy, which has been shown previously.
	
	For the 3-D case, this becomes complicated. The point is that we should do the rescaling on both in the tensor space and the dual tensor space to make sure that the tensor have the same scaling property in both spaces. Although mathematically we may do rescaling only in the tensor space, but if we make our code like this, the tensors will may oscillating and eventually blow up. 
	
	For 3-D cubic tensor network,
	
	\begin{equation}
		||T(\lambda \beta)||\sim ||T(\beta)||^{\lambda}
	\end{equation}
	
	So 
	
	\begin{equation}
		||A(\lambda \beta)||\sim ||A(\beta)||^{\lambda/3}
	\end{equation}
	
	\begin{equation}
		||B(\lambda \beta)||\sim ||B(\beta)||^{2\lambda/3}
	\end{equation}
	
	\begin{equation}
		||C(\lambda \beta)||\sim ||C(\beta)||^{\lambda/3}
	\end{equation}
	
	Then contract over a cube with 8 $C$ tensors 
	
	\begin{equation}
		||S'_s(\lambda \beta)||\sim|| S_s(\beta)||^{8\lambda/3}
	\end{equation}
	
	to keep linearity, we should rescale by $S^{5/3}$
	
	\begin{equation}
		||S(\lambda \beta)||\sim ||S(\beta)||^{\lambda}
	\end{equation}
	
	A practical choice of the factor could be based on the first SVD step, if we denote the square root of the largest singular value of the SVD of $B$ by $f_1$. We know that $f_1\sim S^{1/3}$. So the first scaling factor would be
	\begin{equation}
		f_1^5
	\end{equation}
	
	Then after SVD, we have
	
	\begin{equation}
		||D(\lambda \beta)||\sim ||D(\beta)||^{\lambda/2}
	\end{equation}
	
	Contract over an octahedron, which have 6 $D$ tensors.
	
	\begin{equation}
		||T'_s(\lambda \beta)||\sim|| T'_s(\beta)||^{3\lambda}
	\end{equation}
	
	So we need to divide $T'_s$  by $T^2$. Similarly if we denote the square root of the largest singular value of the SVD of $S$ by $f_2$, since $f_2\sim T^{1/2}$, we need to rescale by 
	\begin{equation}
		f_2^4
	\end{equation}
	
	Then 
	
	\begin{equation}
		||T'(\lambda \beta)||\sim ||T'(\beta)||^{\lambda}
	\end{equation}
	
	Our overall rescaling factor $f$  for the cubic Ising tensor network would be
	
	\begin{equation}
		f=(f_1^5, f_2^4)=(a,b)
	\end{equation}
	
	Numerically, based on our test on the program, (5,4) are the correct value to make iteration converge. Other powers would either cause the tensor goes to zero, or diverge.The notation $(a,b)$ will be used in the next section.
	
	\subsection{Derivation of free energy}
\label{marker44}
		
	For a cubic tensor network system, we want to calculate the partition function.
	\begin{equation}
		Z=\sum e^{-\beta H(\sigma)}=\sum_{}TTTT...
	\end{equation}
	
	Let's assume we have $N$ tensors ($3N$ spins). After converting to the dual tensor network, we have $\frac{3N}{8}$ of $S$ tensors, then after one step of RG, we have $\frac{N}{8}$ of $T'$ tensors.
	
	\begin{multline}
		Z=\sum \prod T=a_1^{\frac{3N}{8}} \sum \prod_{\frac{3N}{8}} S=a_1^{\frac{3N}{8}}b_1^{\frac{N}{8}} \sum \prod_{\frac{N}{8}} T'
		=\\
		a_1^{\frac{3N}{8}}b_1^{\frac{N}{8}}a_2^{\frac{3N}{8^2}}  \sum \prod_{\frac{3N}{8^2}} S'=a_1^{\frac{3N}{8}}b_1^{\frac{N}{8}}a_2^{\frac{3N}{8^2}} b_2^{\frac{N}{8^2}} \sum \prod_{\frac{N}{8^2}} T''=... 
	\end{multline} 
	
	In the end, if we have a very large finite $N$, we'll get
	
	\begin{equation}
		Z=a_1^{\frac{3N}{8}}b_1^{\frac{N}{8}}  a_2^{\frac{3N}{8^2}}b_2^{\frac{N}{8^2}} ...a_n^{\frac{3N}{8^{n}}}b_n^{\frac{N}{8^{n}}}.....Tr(T^*)
	\end{equation} 
	
	The free energy per unit spin can be calculated as
	
	\begin{multline}
		F=-\frac{1}{(3N)\beta}ln Z=-\frac{1}{(3)\beta}(\frac{3}{8}ln a_1+\frac{1}{8}ln b_1+\\\frac{3}{8^2}ln a_2+\frac{1}{8^2}ln b_2+....\frac{3}{8^{n}}ln a_n+\frac{1}{8^{n}}ln b_n+...\frac{ln Tr(T^*)}{N})
	\end{multline}

	The factor of 3 comes from the fact that each $T$ corresponds to 3 spins.  If  $N$ goes to infinity,
	
	\begin{multline}
		F=-\frac{1}{(3)\beta}ln Z=-\frac{1}{(3)\beta}(\frac{3}{8}ln a_1+\frac{1}{8}ln b_1+\\\frac{3}{8^2}ln a_2+\frac{1}{8^2}ln b_2+....\frac{3}{8^{n}}ln a_n+\frac{1}{8^{n}}ln b_n...)
	\end{multline}
	
	Here $a_{i}(b_i)$ is a function of temperature and magnetic field, so $a_i=a_i(T,h)$,$b_i=b_i(T,h)$. So we have a function of the free energy as function of $h$ and $T$, so we can calculate physical quantities based on the free energy. These results are similar to 2-D.

	Practically, the $f$ factor converges very fast. So we actually take $n=4$ and the free energy is approximated by
		\begin{multline}
		F=-\frac{1}{(3)\beta}ln Z=-\frac{1}{(3)\beta}(\frac{3}{8}ln a_1+\frac{3}{8^2}ln a_2+\frac{3}{8^3}ln a_3+\\\frac{3}{8^{3}\times7}ln a+\frac{1}{8}ln b_1+\frac{1}{8^2}ln b_2+\frac{1}{8^3}ln b_3+\frac{1}{8^{3}\times7}ln b)
	\end{multline}
 
	\subsection{Numerical results} 	
	We do the calculation for a $32\times 32\times 32$ cubic tensor network which corresponds to 5 TRG steps. We also compare this tensor results with the Monte Carlo results. The calculation cost scales with the number of sites for the Monte Carlo method, while for TRG, it scales with the number of iterations. The contraction part of TRG takes most of the computational time. The  tensor contractor NCON\cite{pfeifer2014ncon} is used to carry out this part.
	
\begin{figure}[!htbp]
		\begin{center}
			\includegraphics[width=100mm]{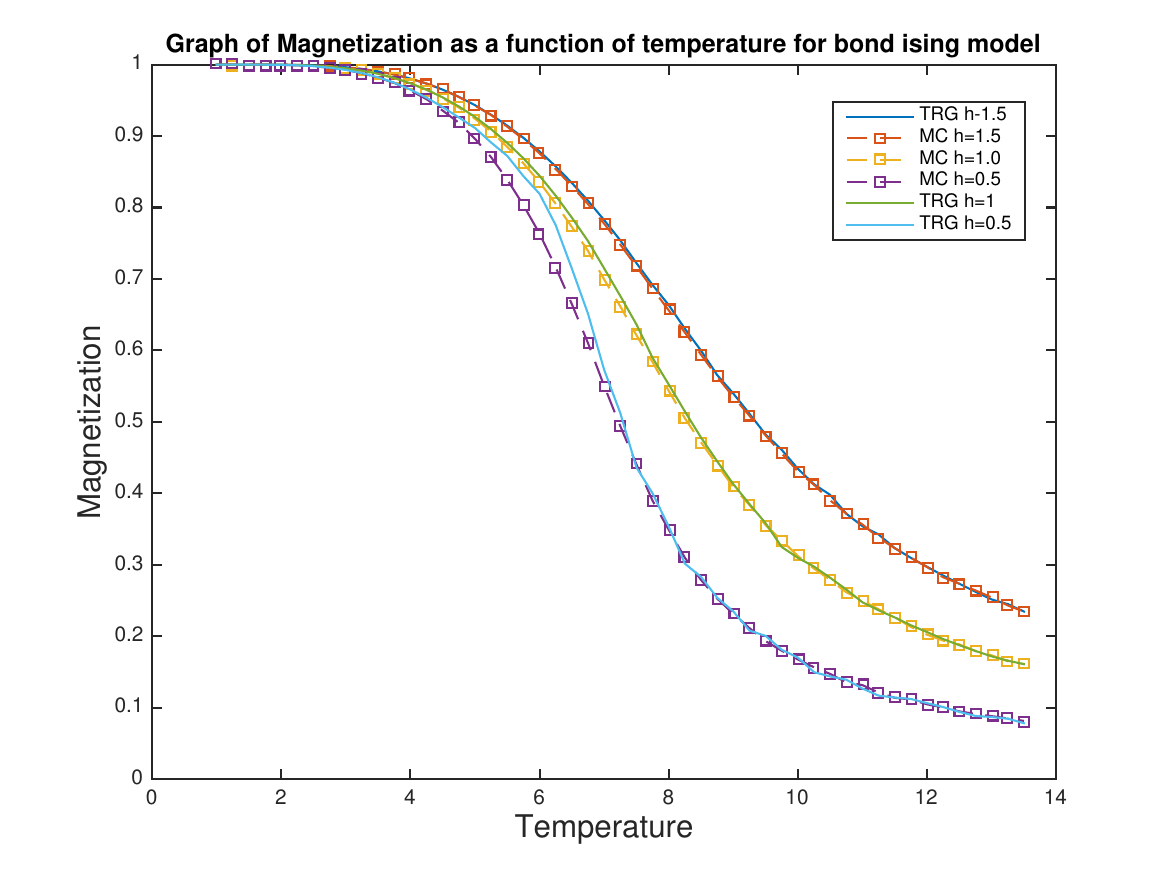}
			\caption{Magnetization as a function of temperature at different $h$. When the magnetic field gets higher, the truncation error gets lower, therefore the results agree better. }
			\label{hall}
		\end{center}
	\end{figure}

\begin{figure}[!htbp]
		\begin{center}
			\includegraphics[width=100mm]{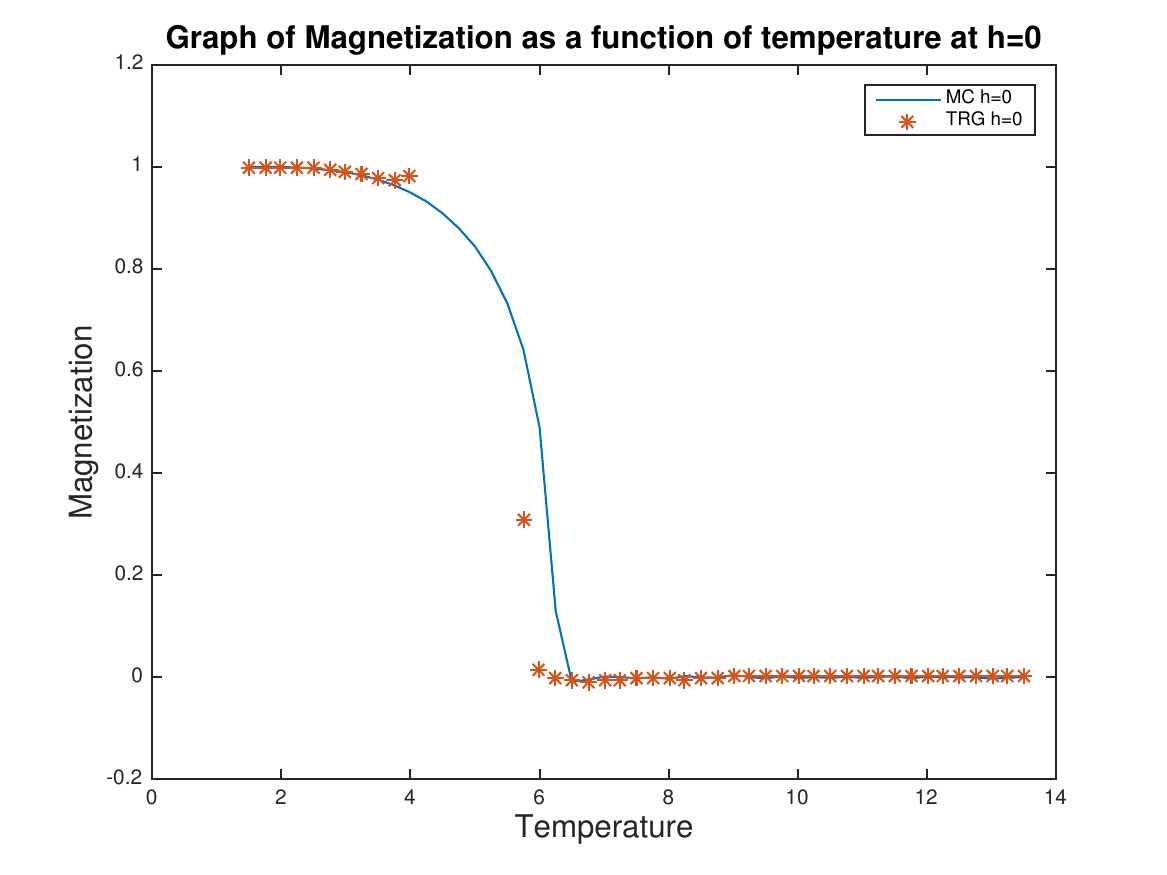}
			\caption{Magnetization as a function of temperature at $h=0$. We notice that for overall bond dimension $D=2$, 3D-TRG blows up at around criticality. The initial condition for CPD iteration is set at (1, 0.5, 0.5, 2) using Tensor Toolbox. The CPD is carried out by consecutive best rank-1 approximation to achieve stability. At criticality, we see a peak, which is not accurate, therefore it is not drawn. }
			\label{hall0}
		\end{center}
	\end{figure}

	The CPD values are kept at 2. The alternate least square method is used to conduct the decomposition. The comparison is carry out at $h=0$, $h=0.5$, $h=1.0$ and $h=1.5$.
	See Figure \ref{hall} and Figure \ref{hall0}.  The curved lines correspond to the TRG results and the dotted lines correspond to the Monte Carlo results. Monte Carlo results are carried out using the Metropolis algorithms. 
	
	\FloatBarrier
	
	In these figures, the initial tensor bond dimension starts at 2.   The bond dimension for CPD is kept at 2 (where 16 is about the theoretical rank).  4 singular values (out of 8) are kept for the first SVD step and 2 singular values (out of 16) are kept at the second SVD step in order for the renormalized tensor to have the same bond dimension. 

    We should notice that our current numerical results agrees well with the Monte Carlo results at large magnetic field. The reason is that, at a large $h$, the largest singular value or CPD value carries most information of the tensor. Therefore at a high $h$, even if we keep one CPD value, we will be able to accurately calculate the magnetization. When $h=0$, due to the symmetry, the singular values spread out, therefore larger bond dimensions are needed.
    
    We should notice that the currently bond dimension is not accurate at around criticality. The singular value spread out at criticality under the current bond dimension. Using the numerical differentiation, we will have a small free energy at zero magnetic field, therefore it will lead to a large magnetization. At criticality, under current bond dimension, what we will see is a peak for magnetization, which is a result of the previous reason. This peak decreases when the bond dimension is increased. But would not disappear under our computational ability.

	\subsection{Current restrictions of 3D-TRG method} 
	The most time consuming part comes from tensor contraction from tensor $C$ to tensor $S$.  We are summing over a cube with 12 edges and 8 external legs. On a PC, NCON can deal with bond dimension 6, while a direct for loop contraction can only deal with 2 bond dimensions. The accuracy depends greatly on how well can the truncation error be controlled. Assuming we keeping the same bond dimension for all the tensors,Πthe time is proportional to $D^{20}$, $D$ is the bond dimension. Therefore we would say that model with larger bond dimension is not currently verified and there are many possibilities, which need further developments. We do have to be careful that when the bond dimension gets large, there truncation error may still gets bigger, due to the fact that higher dimension tensor have more legs, therefore increase the truncation error.

	Another potential restriction is that the CPD optimization problem is not convex, therefore a initial condition have to be carefully chosen.

	Although a systematical application of this method depends greatly on the details and mathematical understandings of CPD, the right algorithms to use and the computational ability of computers, we should not under estimate the importance of tensor network method, since SVD is well-controlled. CPD could also be well-controlled, and it is a non-perturbative method and may be applied to any systems in any dimensions, since our framework is exact without truncation.
	
	\section{Tensor RG for higher dimensional tensor network}\label{4d}
	
	We consider an n-dimensional tensor network, which can be mapped to a n-D Ising model. Knowledge of \textbf{regular polytopes} is needed\cite{coxeter1973regular}. Polytopes are the higher dimensional generalization of polyhedrons and polygons.  In mathematics, \textbf{honeycomb} means close packing of the polytopes.
	
	This method can be generalized to any higher dimension. For an n-cube honeycomb tensor network, we should utilize the topological duality relation between an n-cube and an n-orthoplex.  Especially, in 4D, this duality become the  duality between a 8-cell(tesseract) and 16-cell. 
	
\begin{figure}[!htbp]
		\begin{center}
			\includegraphics[width=80mm]{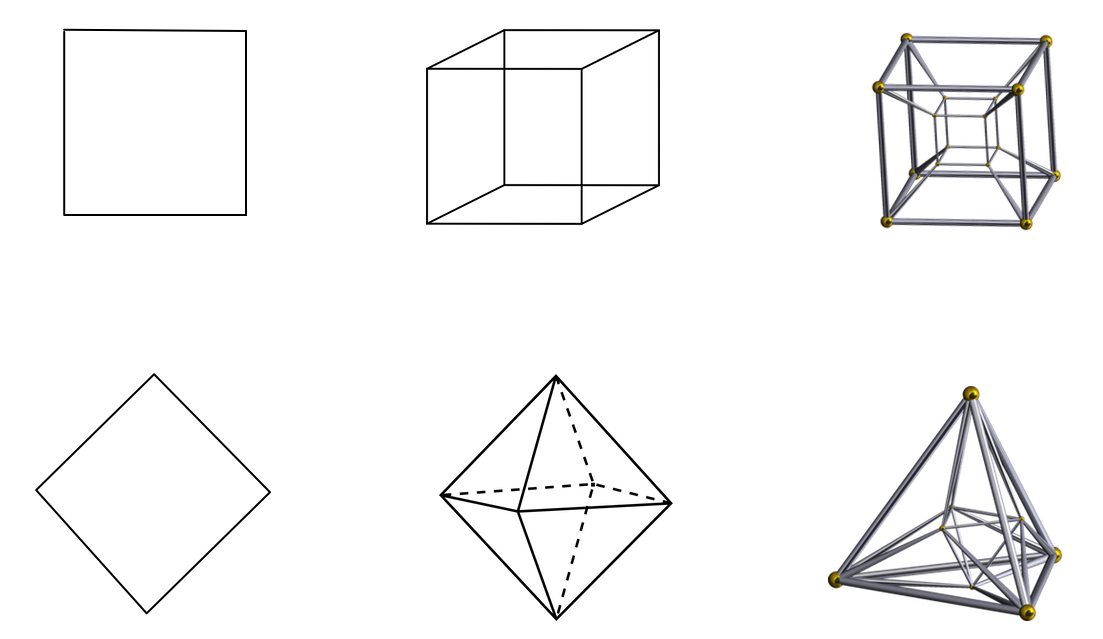}
			\caption{n-cube and n-orthoplex (n=2,3,4), 4D cases credit to Stella Software.}
			\label{dual}
		\end{center}
	\end{figure}

	Figure \ref{dual} shows this topological duality relation in 2, 3, and 4 dimensions.
	
	We can see that the in 2-D, 2-cube and 2-orthoplex are all square. This observation provides a different view of  TRG in 2-D, i.e, we are still transforming into the dual space. The interesting part is: the dual space is the same as the original space.

	The process of RG should be understood within the framework of the truncation sequence from an n-cube to an n-orthoplex. For example, in 4D, we are going from the original space to its dual space by going through the sequence of 4-cube (8-cell), truncated 4-cube, rectified 4-cube, bitruncated 4-cube, (16-cell). In the language of tensor network, the truncation of an n-cube corresponds to the CPD of an n-way tensor (2n-way before grouping indices), rectification of an n-cube corresponds to sum over the edges, then an SVD is needed to go from the rectified n-cube to a bitruncated n-cube, then after summing over an n-cube tensor, we get  the 16-cell dual space.
	
	In order to dual back to the original space, we have to do another SVD on the n-orthoplex tensor network, then contact over the n-orthoplex, we get back to the n-cube tensor network. 
	
	Scaling is also needed in n-D, and corresponding factor can be calculated based on the requirement of converging to a finite value at infinity. Then we can follow the same procedure and calculate the free energy.

	We should notice that spacial dimension 4 has different tessellation property than 3-D, and it is similar to 2-D. In 2-D we have 3 different kinds of regular tessellation: triangle, square, and hexagon. In 3-D, however cubic honeycomb is the only regular tessellation. In 4-D, things get complicated again. We have 8-cell (4-cube) tessellation, 16-cell tessellation and 24-cell tessellation.   In 5-D or higher dimensions cubic honeycomb is the only regular honeycomb that can tessellate the entire space.
	
	In 4-D, 120-cell and 600-cell also truncates to each other but they don't form any regular honeycomb. 4-simplex (5-cell) truncates to itself but it doesn't form any honeycomb.
	
	In all, our framework can applied to any regular tessellated Euclidean tensor network (tensor network that can be represented as the repeating of one regular all-space-filling polytope). 
	
The tensor renormalization group process in any dimension for a regular tessellated network can be illustrated using the truncation sequence of a certain polytope. The truncation process may correspond to the tensor rank decomposition. Converting to the dual tensor network is needed.

	\section{Discussion}\label{discuss}
	\subsection{Applications to quantum system}
	
	Our framework is a geometric way to contract a tensor network. It can be applied to both classical model and quantum model. We can use the Trotter-Suzuki formula to convert $d$ dimensional quantum system to a  $d+1$ dimensional classical system. After the conversion, we can represent the classical model in a solvable tensor network and then do RG on the tensor network to find the thermal-dynamical quantities. Details can be found at\cite{zhao2010renormalization}. 
	
	\subsection{Potential problems of TRG in Higher dimensions}
	
	(1) Truncation error.
	
	For 2-D TRG, we are keeping D values of the $D\times D$ matrices from SVD, assuming $D$ is the bond dimension of the matrix. Previous result tells us that the accuracy, of the TRG increases with the bond dimension $D$. For a 3-D cubic tensor network, we are actually keeping $D$ values of a $D^4\times D^4$ matrix ($D^8$ tensor). For a n-D cubic tensor network, we are keeping $D$ values of a $D^{n-1}\times D^{n-1}$ matrices. It's natural that the truncation error grows with the dimension, therefore the accuracy of TRG gets lower for higher dimensional system.
	
	(2) Calculation cost.
	
	For TRG in higher dimensional system, we have to sum over an n-cube and an n-orthoplex. The costs of this summation grows exponentially with the spacial dimension. Programming is also significantly difficult for a higher dimensional system. Therefore Monte Carlo method is still a good choice for a classical systems. For quantum system, due to the sign problem, TRG looks like a good candidate, but further research still need to be done in order to test the accuracy of this method.
		
	\subsection{Is CPD the only choice?}

	Our starting point of this framework is the tessellation geometry, therefore we should keep the same number of tensor legs as the truncated polytope has. Careful readers may notice that the key equation for a truncation is, (same as equation \ref{25}).
		\begin{equation}
		T_{mnp}\approx \sum A'_{\mu\nu m} A'_{\nu \rho n} A'_{\rho \mu p}
		\end{equation} 
		
	CPD can be regarded as the easiest way to implement this equation, since for CPD $\mu=\nu=\rho$, and the rank is defined as the smallest number of sums. In general HOSVD(Tucker decomposition) also satisfies this equation, since we can rewrite our core tensor of HOSVD as a vector. This equation is not restricted to these two cases, since we may have $\mu\ne \nu\ne\rho$. For 2-D, this equation reduce to SVD, since we have a matrix to decompose and the corner of the truncation is a edge, while for 3-D it is a face.
	
	This equation can be regarded as an inverse of a tensor contraction, therefore it is very likely that we can have a better way to do this decomposition. It will be interesting to think about other possibilities that is better that CPD. We notices that this direction had be explored by the papers of S-J Ran\cite{ran2013theory}.

	\subsection{CPD and best rank-r approximation}	
	
	Notice that theoretically we are using CPD as our decomposition method, although numerically we need to pick a fixed number as our rank. This is generally termed as the best rank-r approximation. For 2-D, SVD represents the best rank-r approximation.  Currently, there is no general way to find the exact rank of an arbitrary  n-way tensor. Therefore, practically, people use the best rank-r approximation as the numerical approximation of CPD. Notice that the tensor rank problem is ill-posed since there exists tensors that has a fixed theoretical rank but it can be approximated arbitrarily close by a numerical rank that is lower that theoretical rank, see details in\cite{illposed}.
	
	\section{Conclusions}
	
	In this article, we discussed the generalization of TRG to higher dimensions and a systematical contraction scheme is proposed. This method currently agrees well with the Monte Carlo results at a high magnetic field. Further development in CPD is needed to give rise to a more accurate physical result for any magnetic fields.
	
	\section{Acknowledgement}
	
	Great thanks should be given to one referee of this paper who gives many insightful and helpful suggestions, and Dr. Yuan-Ming Lu for his discussions and comments. I also want to give my sincere thanks to the Ohio State University Physics Department who financially supported my study. This research did not receive any specific grant from funding agencies in the public, commercial, or not-for-profit sectors.

\section*{References}	

\bibliography{mybibfile}

\begin{thebibliography}{10}
\expandafter\ifx\csname url\endcsname\relax
  \def\url#1{\texttt{#1}}\fi
\expandafter\ifx\csname urlprefix\endcsname\relax\def\urlprefix{URL }\fi
\expandafter\ifx\csname href\endcsname\relax
  \def\href#1#2{#2} \def\path#1{#1}\fi

\bibitem{fannes1992finitely}
M.~Fannes, B.~Nachtergaele, R.~F. Werner,
  \href{http://dx.doi.org/10.1007/BF02099178}{Finitely correlated states on
  quantum spin chains}, Communications in Mathematical Physics 144~(3) (1992)
  443--490.
\newblock \href {http://dx.doi.org/10.1007/BF02099178}
  {\path{doi:10.1007/BF02099178}}.
\newline\urlprefix\url{http://dx.doi.org/10.1007/BF02099178}

\bibitem{ostlund1995thermodynamic}
S.~\"Ostlund, S.~Rommer,
  \href{http://link.aps.org/doi/10.1103/PhysRevLett.75.3537}{Thermodynamic
  limit of density matrix renormalization}, Phys. Rev. Lett. 75 (1995)
  3537--3540.
\newblock \href {http://dx.doi.org/10.1103/PhysRevLett.75.3537}
  {\path{doi:10.1103/PhysRevLett.75.3537}}.
\newline\urlprefix\url{http://link.aps.org/doi/10.1103/PhysRevLett.75.3537}

\bibitem{verstraete2004renormalization}
F.~Verstraete, J.~I. Cirac, Renormalization algorithms for quantum-many body
  systems in two and higher dimensions, arXiv preprint cond-mat/0407066.

\bibitem{white1992density}
S.~R. White, \href{http://link.aps.org/doi/10.1103/PhysRevLett.69.2863}{Density
  matrix formulation for quantum renormalization groups}, Phys. Rev. Lett. 69
  (1992) 2863--2866.
\newblock \href {http://dx.doi.org/10.1103/PhysRevLett.69.2863}
  {\path{doi:10.1103/PhysRevLett.69.2863}}.
\newline\urlprefix\url{http://link.aps.org/doi/10.1103/PhysRevLett.69.2863}

\bibitem{vidal2007entanglement}
G.~Vidal,
  \href{http://link.aps.org/doi/10.1103/PhysRevLett.99.220405}{Entanglement
  renormalization}, Phys. Rev. Lett. 99 (2007) 220405.
\newblock \href {http://dx.doi.org/10.1103/PhysRevLett.99.220405}
  {\path{doi:10.1103/PhysRevLett.99.220405}}.
\newline\urlprefix\url{http://link.aps.org/doi/10.1103/PhysRevLett.99.220405}

\bibitem{levin2007tensor}
M.~Levin, C.~P. Nave,
  \href{http://link.aps.org/doi/10.1103/PhysRevLett.99.120601}{Tensor
  renormalization group approach to two-dimensional classical lattice models},
  Phys. Rev. Lett. 99 (2007) 120601.
\newblock \href {http://dx.doi.org/10.1103/PhysRevLett.99.120601}
  {\path{doi:10.1103/PhysRevLett.99.120601}}.
\newline\urlprefix\url{http://link.aps.org/doi/10.1103/PhysRevLett.99.120601}

\bibitem{kadanoff1966spin}
L.~P. Kadanoff, \href{http://dx.doi.org/10.1007/BF02710808}{Spin-spin
  correlations in the two-dimensional ising model}, Il Nuovo Cimento B
  (1965-1970) 44~(2) (1966) 276--305.
\newblock \href {http://dx.doi.org/10.1007/BF02710808}
  {\path{doi:10.1007/BF02710808}}.
\newline\urlprefix\url{http://dx.doi.org/10.1007/BF02710808}

\bibitem{zhao2010renormalization}
H.~H. Zhao, Z.~Y. Xie, Q.~N. Chen, Z.~C. Wei, J.~W. Cai, T.~Xiang,
  \href{http://link.aps.org/doi/10.1103/PhysRevB.81.174411}{Renormalization of
  tensor-network states}, Phys. Rev. B 81 (2010) 174411.
\newblock \href {http://dx.doi.org/10.1103/PhysRevB.81.174411}
  {\path{doi:10.1103/PhysRevB.81.174411}}.
\newline\urlprefix\url{http://link.aps.org/doi/10.1103/PhysRevB.81.174411}

\bibitem{TNR}
G.~Evenbly, G.~Vidal,
  \href{http://link.aps.org/doi/10.1103/PhysRevLett.115.180405}{Tensor network
  renormalization}, Phys. Rev. Lett. 115 (2015) 180405.
\newblock \href {http://dx.doi.org/10.1103/PhysRevLett.115.180405}
  {\path{doi:10.1103/PhysRevLett.115.180405}}.
\newline\urlprefix\url{http://link.aps.org/doi/10.1103/PhysRevLett.115.180405}

\bibitem{xie2012coarse}
Z.~Y. Xie, J.~Chen, M.~P. Qin, J.~W. Zhu, L.~P. Yang, T.~Xiang,
  \href{http://link.aps.org/doi/10.1103/PhysRevB.86.045139}{Coarse-graining
  renormalization by higher-order singular value decomposition}, Phys. Rev. B
  86 (2012) 045139.
\newblock \href {http://dx.doi.org/10.1103/PhysRevB.86.045139}
  {\path{doi:10.1103/PhysRevB.86.045139}}.
\newline\urlprefix\url{http://link.aps.org/doi/10.1103/PhysRevB.86.045139}

\bibitem{jiang2008accurate}
H.~C. Jiang, Z.~Y. Weng, T.~Xiang,
  \href{http://link.aps.org/doi/10.1103/PhysRevLett.101.090603}{Accurate
  determination of tensor network state of quantum lattice models in two
  dimensions}, Phys. Rev. Lett. 101 (2008) 090603.
\newblock \href {http://dx.doi.org/10.1103/PhysRevLett.101.090603}
  {\path{doi:10.1103/PhysRevLett.101.090603}}.
\newline\urlprefix\url{http://link.aps.org/doi/10.1103/PhysRevLett.101.090603}

\bibitem{ran2013theory}
S.-J. Ran, B.~Xi, T.~Liu, G.~Su,
  \href{http://link.aps.org/doi/10.1103/PhysRevB.88.064407}{Theory of network
  contractor dynamics for exploring thermodynamic properties of two-dimensional
  quantum lattice models}, Phys. Rev. B 88 (2013) 064407.
\newblock \href {http://dx.doi.org/10.1103/PhysRevB.88.064407}
  {\path{doi:10.1103/PhysRevB.88.064407}}.
\newline\urlprefix\url{http://link.aps.org/doi/10.1103/PhysRevB.88.064407}

\bibitem{nishino2001two}
T.~Nishino, Y.~Hieida, K.~Okunishi, N.~Maeshima, Y.~Akutsu, A.~Gendiar,
  \href{http://ptp.oxfordjournals.org/content/105/3/409.abstract}{Two-dimensional
  tensor product variational formulation}, Progress of Theoretical Physics
  105~(3) (2001) 409--417.
\newblock \href {http://dx.doi.org/10.1143/PTP.105.409}
  {\path{doi:10.1143/PTP.105.409}}.
\newline\urlprefix\url{http://ptp.oxfordjournals.org/content/105/3/409.abstract}

\bibitem{garcia2013renormalization}
A.~Garc\'{\i}a-S\'aez, J.~I. Latorre,
  \href{http://link.aps.org/doi/10.1103/PhysRevB.87.085130}{Renormalization
  group contraction of tensor networks in three dimensions}, Phys. Rev. B 87
  (2013) 085130.
\newblock \href {http://dx.doi.org/10.1103/PhysRevB.87.085130}
  {\path{doi:10.1103/PhysRevB.87.085130}}.
\newline\urlprefix\url{http://link.aps.org/doi/10.1103/PhysRevB.87.085130}

\bibitem{li2011linearized}
W.~Li, S.-J. Ran, S.-S. Gong, Y.~Zhao, B.~Xi, F.~Ye, G.~Su,
  \href{http://link.aps.org/doi/10.1103/PhysRevLett.106.127202}{Linearized
  tensor renormalization group algorithm for the calculation of thermodynamic
  properties of quantum lattice models}, Phys. Rev. Lett. 106 (2011) 127202.
\newblock \href {http://dx.doi.org/10.1103/PhysRevLett.106.127202}
  {\path{doi:10.1103/PhysRevLett.106.127202}}.
\newline\urlprefix\url{http://link.aps.org/doi/10.1103/PhysRevLett.106.127202}

\bibitem{wenninger2003dual}
M.~J. Wenninger, Dual models, Cambridge University Press, 2003.

\bibitem{kolda2009tensor}
T.~G. Kolda, B.~W. Bader, \href{http://dx.doi.org/10.1137/07070111X}{Tensor
  decompositions and applications}, SIAM Review 51~(3) (2009) 455--500.
\newblock \href {http://dx.doi.org/10.1137/07070111X}
  {\path{doi:10.1137/07070111X}}.
\newline\urlprefix\url{http://dx.doi.org/10.1137/07070111X}

\bibitem{kruskal1989rank}
J.~B. Kruskal, \href{http://dl.acm.org/citation.cfm?id=120565.120567}{Multiway
  data analysis} (1989) 7--18.
\newline\urlprefix\url{http://dl.acm.org/citation.cfm?id=120565.120567}

\bibitem{bader2012matlab}
B.~W. Bader, T.~G. Kolda, et~al., Matlab tensor toolbox version 2.6, Available
  online, January 7.

\bibitem{acar2011scalable}
E.~Acar, D.~M. Dunlavy, T.~G. Kolda,
  \href{http://dx.doi.org/10.1002/cem.1335}{A scalable optimization approach
  for fitting canonical tensor decompositions}, Journal of Chemometrics 25~(2)
  (2011) 67--86.
\newblock \href {http://dx.doi.org/10.1002/cem.1335}
  {\path{doi:10.1002/cem.1335}}.
\newline\urlprefix\url{http://dx.doi.org/10.1002/cem.1335}

\bibitem{pfeifer2014ncon}
R.~N. Pfeifer, G.~Evenbly, S.~Singh, G.~Vidal, Ncon: A tensor network
  contractor for matlab, arXiv preprint arXiv:1402.0939.

\bibitem{coxeter1973regular}
H.~S.~M. Coxeter, Regular polytopes, Courier Corporation, 1973.

\bibitem{illposed}
V.~de~Silva, L.-H. Lim, \href{http://dx.doi.org/10.1137/06066518X}{Tensor rank
  and the ill-posedness of the best low-rank approximation problem}, SIAM
  Journal on Matrix Analysis and Applications 30~(3) (2008) 1084--1127.
\newblock \href {http://dx.doi.org/10.1137/06066518X}
  {\path{doi:10.1137/06066518X}}.
\newline\urlprefix\url{http://dx.doi.org/10.1137/06066518X}

\end{thebibliography}

\end{document}